\documentclass[fleqn,usenatbib]{mnras}
\usepackage[T1]{fontenc}
\usepackage{ae,aecompl}
\usepackage{graphicx,color}
\usepackage{amsmath}
\usepackage{amssymb}
\usepackage{ascmac}
\usepackage{bm,braket}
\usepackage{grffile}
\usepackage{fancybox}

\numberwithin{equation}{section}

\newcommand{\DD}{D_{\rm\scriptscriptstyle X}D_{\rm\scriptscriptstyle Y}}
\newcommand{\RX}{R_{\rm\scriptscriptstyle X}}
\newcommand{\RY}{R_{\rm\scriptscriptstyle Y}}
\newcommand{\data}[1]{{\tt data\_H}$_{#1}$}
\expandafter\def\expandafter\UrlBreaks\expandafter{\UrlBreaks
 \do\a\do\b\do\c\do\d\do\e\do\f\do\g\do\h\do\i\do\j%
 \do\k\do\l\do\m\do\n\do\o\do\p\do\q\do\r\do\s\do\t%
 \do\u\do\v\do\w\do\x\do\y\do\z\do\A\do\B\do\C\do\D%
 \do\E\do\F\do\G\do\H\do\I\do\J\do\K\do\L\do\M\do\N%
 \do\O\do\P\do\Q\do\R\do\S\do\T\do\U\do\V\do\W\do\X%
 \do\Y\do\Z}
\title[Relativistic effects at quasi-linear scales]{
Modelling the asymmetry of the halo cross-correlation function with relativistic effects at quasi-linear scales
}
\author[S. Saga et al.]
{Shohei Saga,$^{1}$
Atsushi Taruya,$^{1,2}$
Michel-Andr{\`e}s Breton,$^{3}$
Yann Rasera$^{4}$
\\
$^{1}$Center for Gravitational Physics, Yukawa Institute for Theoretical Physics, Kyoto University, Kyoto 606-8502, Japan\\
$^{2}$Kavli Institute for the Physics and Mathematics of the Universe (WPI), The University of Tokyo Institutes for Advanced Study, \\
The University of Tokyo, 5-1-5 Kashiwanoha, Kashiwa, Chiba 277-8583, Japan \\
$^{3}$Aix Marseille Univ, CNRS, CNES, LAM, Marseille, France\\
$^{4}$Laboratoire Univers et Th{\'e}ories, Observatoire de Paris, Universit{\'e} PSL, CNRS, Universit{\'e} de Paris,\\
5 place Jules Janssen 92190 Meudon, France \\
}
\thisfancyput(14.5cm,0.5cm){\large{YITP-20-40}}
\date{Accepted XXX. Received YYY; in original form ZZZ}
\pubyear{2020}
\begin{document}
\label{firstpage}
\pagerange{\pageref{firstpage}--\pageref{lastpage}}
\maketitle
\begin{abstract}
The observed galaxy distribution via galaxy redshift surveys appears distorted due to redshift-space distortions (RSD).
While one dominant contribution to RSD comes from the Doppler effect induced by the peculiar velocity of galaxies, the relativistic effects, including the gravitational redshift effect, are recently recognized to give small but important contributions.
Such contributions lead to an asymmetric galaxy clustering along the line of sight, and produce non-vanishing odd multipoles when cross-correlating between different biased objects.
However, non-zero odd multipoles are also generated by the Doppler effect beyond the distant-observer approximation, known as the wide-angle effect, and at quasi-linear scales, the interplay between wide-angle and relativistic effects becomes significant.
In this paper, based on the formalism developed by Taruya et al., we present a quasi-linear model of the cross-correlation function taking a proper account of both the wide-angle and gravitational redshift effects, as one of the major relativistic effects.
Our quasi-linear predictions of the dipole agree well with simulations even at the scales below $20\,h^{-1}\,$Mpc, where non-perturbative contributions from the halo potential play an important role, flipping the sign of the dipole amplitude. 
When increasing the bias difference and redshift, the scale where the sign flip happens is shifted to a larger scale.
We derive a simple approximate formula to quantitatively account for the behaviors of the sign flip.
\end{abstract}

\section{Introduction}

The large-scale structure observed via spectroscopic surveys appears distorted due to various physical processes.
Such distortions are arising mainly from the peculiar velocity of galaxies through the Doppler effect (standard Doppler effect), referred to as the redshift-space distortions (RSD)~\citep{1987MNRAS.227....1K,1992ApJ...385L...5H}.
Measuring the anisotropic galaxy clustering in redshift space offers a unique opportunity to probe the growth of cosmic structure induced by gravity.
In this respect, RSD has been used as a powerful tool for testing gravity on cosmological scales~(e.g.,~\citealt{2008Natur.451..541G,2008APh....29..336L,2009MNRAS.393..297P,2012MNRAS.426.2719R,2013MNRAS.433.1202S,2017MNRAS.470.2617A}).

It has been recently recognized that the observed galaxy distribution is further distorted due to the relativistic corrections that arise from the light propagation in an inhomogeneous Universe~(e.g.,~\citealt{2004MNRAS.348..581P,2009PhRvD..80h3514Y,2010PhRvD..82h3508Y,2011PhRvD..84f3505B,2011PhRvD..84d3516C,2014CQGra..31w4001Y}, and see also \citealt{1987MNRAS.228..653S} for pioneer work).
For instance, on top of the Doppler effect by the peculiar velocity of galaxies, the observed redshift of galaxies is affected by the gravitational potential at their position and along the light path.
The angular position of objects on the sky is also changed by gravitational lensing effect. These relativistic effects lead to an asymmetry in the two-point correlation function and to an imaginary part of power spectrum between different biased objects~\citep{2013MNRAS.434.3008C,2009JCAP...11..026M,2012arXiv1206.5809Y,2014PhRvD..89h3535B,2018JCAP...03..019T}.
Since the asymmetric distortions carry the additional information on the nature of gravity, the measurement of them can offer new possibilities to test gravity on cosmological scales.
For example, \citet{2018JCAP...05..061B} have pointed out that the asymmetric distortion can serve as a test of Einstein's equivalence principle on cosmological scales.
Observationally, \citet{2017MNRAS.470.2822A} have claimed the detection of the asymmetry using SDSS BOSS DR12 CMASS galaxy sample and it is consistent with the gravitational redshift effect predicted by general relativity.

Importantly, the asymmetry produced by the relativistic effects leads to non-vanishing odd multipole moments along the line-of-sight direction.
Thus, if there are no other mechanisms to produce odd multipole moments, their non-zero signals would give a strong indication for relativistic effects.
However, on top of the relativistic effects, the standard Doppler effect beyond the distant-observer limit, referred to as the wide-angle effects, is known to produce the non-vanishing odd multipole moments~(e.g.,~\citealt{1994MNRAS.266..219F,1996ApJ...462...25Z,1996MNRAS.278...73H,1998ApJ...498L...1S,2000ApJ...535....1M,2004ApJ...614...51S,2004ApJ...615..573M,2008MNRAS.389..292P}).
Disentangling the two effects are therefore important, and a quantitative prediction and modelling of them are crucial for future measurements.
Along the line of this, \citet{2019MNRAS.483.2671B} have created halo catalogue (from the raytracing of large $N$-body simulations) taking a proper account of both the wide-angle and all possible relativistic effects at first order in metric perturbation.
Then, they studied the asymmetric halo clustering by measuring the dipole moment of the cross-correlation function between halo populations with different masses.
They found that the dipole moment is dominated by the gravitational redshift effect at small scales, and the wide-angle effect at large scales.
In particular, the non-linear impact of the gravitational redshift effect appears manifest, and the deviation from linear theory is significant at small scales.

The aim of this paper is to develop an analytical model which quantitatively predicts and explains the dipole cross-correlations measured in numerical simulations.
To do so, we need to go beyond linear theory, taking consistently both the wide-angle and gravitational redshift effects into account.
Recently, based on the Zel'dovich approximation (first-order Lagrangian perturbation theory)~\citep{1970A&A.....5...84Z,1989RvMP...61..185S}, we have developed the quasi-linear formalism of cross-correlation function, taking properly the wide-angle effect into account \citep[see also][which partly include the wide-angle effect valid at quasi-linear scales]{2018MNRAS.479..741C}.
In this paper, on the basis of this formalism, we further incorporate the relativistic corrections into the predictions of cross-correlation function \citep[see also][for the prediction based on the third-order Eulerian perturbation theory]{2019JCAP...04..050D}.
Focusing in particular on the gravitational redshift effect as the dominant relativistic correction, we will make a detailed comparison between quasi-linear predictions and $N$-body simulations with both the wide-angle and relativistic effects.
Then we quantitatively discuss the impact of these effects on the cross-correlation function at quasi-linear scales.

This paper is organized as follows. In Sec.~\ref{sec: intro relativistic RSD}, we begin by briefly reviewing the relativistic corrections to the RSD in both linear and non-linear regimes based on the results of $N$-body simulations.
In Sec.~\ref{sec: formulation}, on the basis of \citet{2020MNRAS.491.4162T}, we present a quasi-linear model to analytically compute the cross-correlation function between different biased objects, taking a proper account of both the wide-angle and relativistic effects.
In Sec.~\ref{sec: model prediction}, we present the predictions of the dipole cross-correlation including the wide-angle and relativistic effects.
Sec.~\ref{sec: results} then compare the predictions with $N$-body simulations at various redshifts and for different halo populations.
Finally, Sec.~\ref{sec: summary} is devoted to the summary of new findings and the discussions on the other related works.
In addition, derivation of the analytical expressions in our quasi-linear model as well as the linear theory formulas for cross-correlation function with relativistic effects are presented in Appendix~\ref{appendix: correlation function} and \ref{appendix: linear theory}, respectively, together with supplemental proof that the quasi-linear formalism properly reproduces the linear theory in Appendix~\ref{appendix: linearization}. Also, Appendix~\ref{appendix: halo potential} summarizes the analytic model of the non-perturbative contribution to the gravitational redshift effect.
Appendix~\ref{appendix: b = 0.1} studies the impact of the halo identification in $N$-body simulations on the results of gravitational redshift effects.

Throughout this paper, we assume a flat Lambda cold dark matter ($\Lambda$CDM) model. The fiducial values of cosmological parameters are chosen based on the seven-year WMAP results~\citep{2011ApJS..192...18K}:
$\Omega_{\rm m0} = 0.25733$, $\Omega_{\rm b0} = 0.04356$, $\Omega_{\Lambda0} = 0.74259$, and $\Omega_{\rm r0} = 8.076 \times 10^{-5}$ for the density parameters for matter, baryon, dark energy with equation-of-state parameter $w=-1$, and radiation, respectively, at the present time.
The other cosmological parameters are chosen as $h = 0.72$, $n_{\rm s} = 0.963$, and $\sigma_{8} = 0.801$ for the Hubble parameter, scalar spectral index, and the root-mean square matter density fluctuations with top-hat filter of radius $8 \; h^{-1}$ Mpc. Throughout the paper, we will work with units of $c=1$. 

\section{Relativistic redshift-space distortion}
\label{sec: intro relativistic RSD}

In this paper, we are interested in modelling and predicting the cross-correlation function between different biased objects, particularly focusing on its dipole moment.
In this section, we review the relativistic effects on the observed large-scale structure.
Based on the results of numerical simulation~\citep{2019MNRAS.483.2671B}, we summarize their important findings and a couple of remarks on the analytical modelling.

\subsection{Relativistic effects on observed large-scale structure}
\label{sec: Rel RSD intro}

In an inhomogeneous universe, the propagation of photons from the moving source to the observer is affected by several special and general relativistic effects. One such effect is the (standard) Doppler effect, and solving the geodesic equations with the perturbed Friedmann-Lema{\^i}tre-Robertson-Walker (FLRW) metric, one can deduce how the observed galaxy positions are shifted or distorted by the relativistic effects.
As a result, the observed redshift of the galaxies, determined by the ratio of the photon energy in the rest frame of the source to that measured by the observer, differs generally from the cosmological redshift in the presence of the redshift perturbations along the line-of-sight trajectory. Also, the angular positions are changed by the perturbation of the photon trajectory, known as the gravitational lensing effect. On top of these effects, there is further volume perturbation in which the volume element inferred from the redshift and solid angle is perturbed by the relativistic effects~(e.g.,~\citealt{2011PhRvD..84f3505B,2014CQGra..31w4001Y}), though its impact is shown to be small.
These relativistic effects produce additional distortions in the observed galaxy number count.

In analogy to the cosmic microwave background observations, the redshift perturbations are known to give several effects in the observed large-scale structure \citep{2019MNRAS.483.2671B}. These include the transverse Doppler effect which is proportional to the square of the peculiar velocity, gravitational redshift or Sachs-Wolfe effect, integrated Sachs-Wolfe effect, Shapiro time-delay effect, and light-cone effect.
The light-cone effect comes from the fact that the position of source at the time of interest differs from the position on the light cone due to their peculiar velocity (see, e.g.,~\citealt{2014PhRvD..89h3535B}).

Summing up all the contributions, the observed source position in comoving space, referred to as redshift space, $\bm{s}$, is related to the real-space counterpart, $\bm{x}$, through (e.g.,~\citealt{2011PhRvD..84d3516C}) 
\begin{equation}
\bm{s} = \bm{x}
+ \frac{1}{aH}\left( \bm{v}\cdot \hat{\bm{x}}\right)\hat{\bm{x}} + \bm{\epsilon}(\bm{x}) ~,
\label{eq: s to x}
\end{equation}
where $\hat{\bm{x}}$ is the unit vector defined by $\hat{\bm{x}} = \bm{x}/|\bm{x}|$ and $a$, $H$, and $\bm{v}$ are a scale factor, Hubble parameter, and peculiar velocity of galaxies, respectively.
The second term at the right-hand side corresponds to the (standard) Doppler effect.
In Eq.~(\ref{eq: s to x}), the relativistic corrections coming from the redshift perturbations and lensing effect are fully encapsulated in the quantity $\bm{\epsilon}(\bm{x})$, assuming the weak-field approximation and $|\bm{v}|\ll 1$.  While the redshift perturbation produces a line-of-sight dependent contribution to $\bm{\epsilon}$, the lensing effect induces the contribution perpendicular to the source direction.
The explicit form of $\bm{\epsilon}(\bm{x})$ can be found in the  literature~(e.g.,~\citealt{2010PhRvD..82h3508Y,2011PhRvD..84d3516C,2011PhRvD..84f3505B}).

\begin{figure}
\centering
\includegraphics[width=0.4\columnwidth]{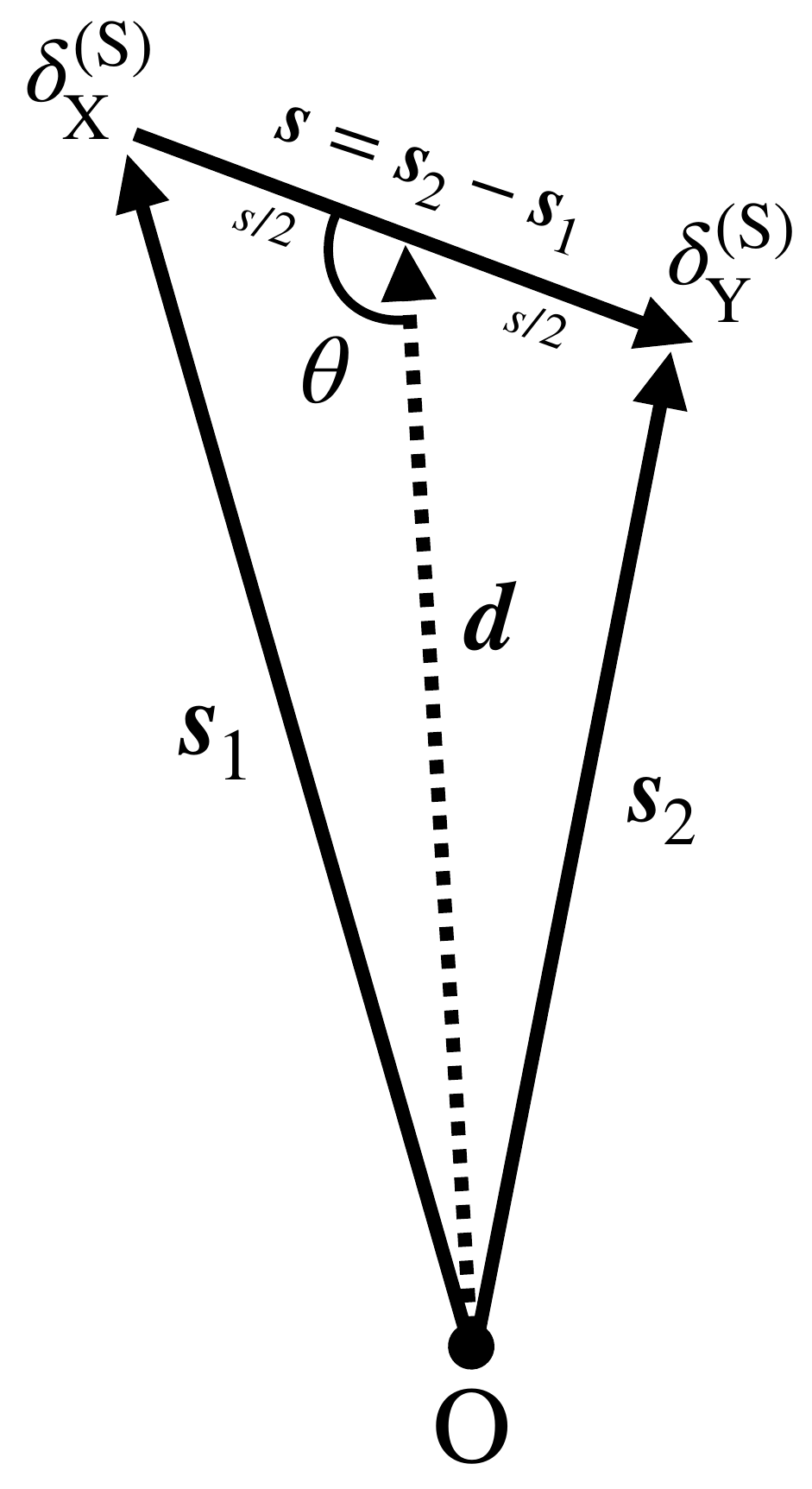}
\caption{Geometric configuration of the cross-correlation function in redshift space using the mid-point line-of-sight definition.
Along the line-of-sight direction $\bm{d} = (\bm{s}_{1}+\bm{s}_{2})/2$, a pair of objects X and Y is found at $\bm{s}_{1}$ and $\bm{s}_{2}$ with respect to the observer~(O), where the density fields, denoted by $\delta^{(\rm S)}_{\rm X}$ and $\delta^{(\rm S)}_{\rm Y}$, are measured, respectively.
The separation vector is defined by $\bm{s} = \bm{s}_{2} - \bm{s}_{1}$.
Misalignment between $\bm{s}$ and $\bm{d}$ is characterised by the angle $\theta$ or the directional cosine given by $\mu \equiv \cos\theta = \hat{\bm{s}}\cdot\hat{\bm{d}}$.
}
\label{fig: configuration}
\end{figure}

Recently, it has been pointed out that the relativistic effects in $\bm{\epsilon}(\bm{x})$ physically break the symmetry of the two-point correlation function with respect to the exchange of pair of objects when cross-correlating between different biased objects.
To quantify the observed clustering anisotropies, we introduce the multipole expansion of the two-point correlation function.
Consider a pair of objects located at redshift-space positions $\bm{s}_{1}$ and $\bm{s}_{2}$, with their separation vector given by $\bm{s}\equiv \bm{s}_{2}-\bm{s}_{1}$, as shown in Fig.~\ref{fig: configuration}. To be specific, we adopt the mid-point line-of-sight direction defined by the vector $\bm{d} \equiv (\bm{s}_{1} + \bm{s}_{2})/2$\footnote{Note that the definition of line-of-sight direction is not unique, and depending on its choice, spurious odd multipoles can be produced \citep[e.g.,][]{2016JCAP...01..048R,2017JCAP...01..032G} even for the auto-correlation. The mid-point line-of-sight we consider here does not produce such a geometrical effect, thus relevant for our purpose.}. Denoting the two-point correlation function by $\xi^{({\rm S})}$, it is generally written as a function of the separation $s=|\bm{s}|$, line-of-sight distance $d=|\bm{d}|$, and directional cosine between the line-of-sight and separation vectors, given by $\mu= \hat{\bm{s}}\cdot\hat{\bm{d}}$, i.e., $\xi^{({\rm S})}(s,d,\mu)$. The multipole correlation function, $\xi_\ell^{\rm(S)}$, is then defined by averaging the correlation function $\xi^{({\rm S})}$ over the directional cosine, weighting with the Legendre polynomials $\mathcal{P}_{\ell}(\mu)$: 
\begin{equation}
\xi^{({\rm S})}_{\ell}(s,d)
= \frac{2\ell+1}{2}\int^{1}_{-1}{\rm d}\mu\; \xi^{({\rm S})}(s,d,\mu) \mathcal{P}_{\ell}(\mu)~.
\label{eq: multipole expansion}
\end{equation}
With this definition, the relativistic effects are shown to give non-vanishing odd multipoles, and in particular, the dipole moment ($\ell=1$) is expected to give the largest signal among other odd multipoles \citep[e.g.,][]{2014PhRvD..89h3535B,2018JCAP...03..019T}.
In this respect, the dipole moment can be exploited as a sensitive probe of relativistic effects, and there are numerous theoretical studies toward a direct detection and measurement of relativistic dipole~\citep[e.g.,][]{2016JCAP...08..021B,2017MNRAS.470.2822A,2018JCAP...05..061B}.

However, the relativistic effects are not the only source to produce the non-zero dipole. The Doppler effect is known to also give a non-zero dipole moment. Note that in the distant-observer limit, under which the directions of the two positions, $\hat{\bm{s}}_{1}$ and $\hat{\bm{s}}_{2}$, are assumed to coincide with a specific line-of-sight direction, $\hat{\bm{s}}_{1}\simeq\hat{\bm{s}}_{2}\simeq\hat{\bm{d}}\equiv\hat{\bm{z}}$, the Doppler effect only generates even multipoles.
Beyond this limit (that is $\hat{\bm{s}}_{1}\neq\hat{\bm{s}}_{2}$ and $\hat{\bm{s}}_{1,2}\neq\hat{\bm{d}}$), however, non-zero multipoles are generally produced, known as the wide-angle effect~(e.g.,~\citealt{1996MNRAS.278...73H,1998ApJ...498L...1S,2000ApJ...535....1M,2004ApJ...614...51S}). Though the amplitude of their odd multipoles is supposed to be small, its impact on the small amplitude of relativistic dipole may be significant, and thus deserves further investigation.

\subsection{Relativistic RSD in simulation}

Until very recently, most of the studies on the detectability of relativistic dipole has been made based on linear theory~(e.g., \citealt{2014PhRvD..89h3535B,2018JCAP...03..019T}). According to these, the dipole cross-correlation function arising from the wide-angle Doppler effect is proportional to the difference of the (linear) bias parameters between two objects, $\Delta b = b_{\rm X} - b_{\rm Y}$, and its amplitude is predicted to be $\xi^{(\rm{s})}_{1}/\Delta b \approx 2\times 10^{-4}$ around $s\approx 100\; h^{-1}{\rm Mpc}$ at $z=0.34$~\citep{2014PhRvD..89h3535B,2018JCAP...03..019T}.
While this is expected to give the most dominant contribution to the dipole, it is found that the the gravitational redshift effect gives the second largest contribution, and its amplitude is estimated as $\xi^{(\rm{s})}_{1}/\Delta b \approx 5\times 10^{-5}$, again, at the scale of $s\approx 100\; h^{-1}{\rm Mpc}$ at $z=0.34$.
Here and in what follows, when cross-correlating the objects with different bias parameters, we always take the object having larger (smaller) bias to be the object $X$ ($Y$), according to Fig.~\ref{fig: configuration}. Thus, the bias difference, $\Delta=b_{\rm X}-b_{\rm Y}$, is always positive. Recalling the fact that in linear theory, the dipole cross-correlation is expressed as $\xi_1^{\rm(s)}\propto\Delta b$ in both the Doppler and gravitational redshift effects with a positive coefficient, the predicted amplitude of the dipole is always positive.

Recently, performing the full-sky $N$-body simulations incorporating the relativistic effects, \citet{2019MNRAS.483.2671B} has numerically investigated the asymmetric galaxy clustering beyond linear theory.
The authors used the PM-AMR $N$-body code RAMSES \citep{2002A&A...385..337T} to run a $\Lambda$CDM cosmological simulation with $4096^3$ particles distributed within a cubic box of 2625 Mpc/$h$ length. The simulation is part of the RayGalGroupSims suite of simulations\footnote{\url{https://cosmo.obspm.fr/raygalgroupsims-relativistic-halo-catalogs/}}. They built a halo catalogue by running the pFoF halo finder \citep{2014A&A...564A..13R} on the full-sky low-redshift particle light-cone. The ray-tracing library \textsc{Magrathea} \citep{Reverdy2014PhD} was used to convert comoving positions to apparent positions. This library finds the null geodesics of a weakly perturbed FLRW metric connecting the observer to sources. Using this method, the information on the true comoving position of the haloes (directly from the full-sky light-cone) as well as their observed positions (observed angular position from null geodesics and redshift perturbations) is obtained. To compute the full redshift decomposition at first order in metric perturbations, the peculiar velocities are evaluated from the haloes themselves while the potential (and its derivatives) are estimated on the AMR gravity light-cone and are interpolated from this grid. Through this
technique, the relativistic contributions such as gravitational redshift, integrated Sachs-Wolfe, and gravitational lensing effects are fully incorporated into the halo catalogue. They further explored, term by term, the contributions to the dipole of the halo cross-correlation function, and studied the behaviour of each term in comparison to linear theory for a wide range of scales at redshift $z < 0.5$.

The important findings of their paper are summarized as follows:
\begin{itemize}\setlength{\itemsep}{0.5em}
\item
At all scales of the $N$-body simulations, relativistic effects involving the line-of-sight integration such as the integrated Sachs-Wolfe, Shapiro time-delay, and gravitational lensing effects, are all subdominant contributions to the dipole cross-correlation.
\item
At large scales $s\gtrsim 30 \; h{\rm Mpc}$, the standard Doppler effect gives the largest contribution to the dipole, and within the statistical error, the measured dipole is in good agreement with linear theory prediction.
Even at $s\lesssim 30\; h{\rm Mpc}$, the predicted dipole for the standard Doppler effect remains reasonable approximation. 
\item
At small scales $s\lesssim 30\; h{\rm Mpc}$, the gravitational redshift effect starts to be dominant, and the amplitude of the dipole cross-correlation sharply drops, flipping the sign from positive to negative. Furthermore, a large deviation from linear theory prediction is found.
\end{itemize}

That is, the gravitational redshift effect is a major source of relativistic contributions to the dipole, and it dominates over other effects at the scales beyond linear regime, where the linear theory prediction fails to reproduce the simulation results.
Hence, in order to quantitatively predict the relativistic dipole, a proper non-linear modelling is essential, taking both the wide-angle and relativistic effects into account. This is our primary focus in the rest of the paper.

\section{Analytic model of relativistic distortions}
\label{sec: formulation}

In this section, extending the quasi-linear formalism of wide-angle RSD developed by~\citet{2020MNRAS.491.4162T}, we construct an analytic model of the cross-correlation function taking the relativistic effects into account.
In what follows, we consider only the gravitational redshift effect, as it gives the most dominant relativistic contribution.
The gravitational redshift effect appears through Eq.~(\ref{eq: s to x}) as a part of the redshift perturbations, written as
\begin{align}
\bm{\epsilon}(\bm{x}) = \epsilon(\bm{x})\hat{\bm{x}}~
 \label{eq: s to x potential}
\end{align}
with the explicit form of $\epsilon(\bm{x})$ given by~(see e.g., \citealt{2011PhRvD..84d3516C})
\begin{equation}
\epsilon(\bm{x}) = -\frac{\phi(\bm{x})}{aH} ~,
\label{eq:epsilon_phi}
\end{equation}
where $\phi(\bm{x})$ is the gravitational potential.

\subsection{Zel'dovich approximation}

Our basis to deal with the cross-correlation function beyond linear theory is the Zel'dovich approximation.
The Zel'dovich approximation~\citep{1969JETP...30..512N,1970A&A.....5...84Z,1989RvMP...61..185S} is the first-order Lagrangian perturbation theory, which allows us to predict the position and motion of mass element, given an initial density field.
For each mass element, the Eulerian position $\bm{x}$ at the time of interest $t$ is related to the Lagrangian coordinate (initial position) $\bm{q}$ through the displacement field, $\bm{\Psi}(\bm{q},t)$. 
In what follows, the objects of our interest to measure the correlation function are assumed to follow the velocity flow of mass distributions.
Then the Eulerian position and velocity of the mass element at $\bm{x}$ are expressed as
\begin{align}
\bm{x}(\bm{q},t) &= \bm{q} + \bm{\Psi}(\bm{q},t) ~, \label{eq: x to q}\\
\bm{v}(\bm{x}) &=a \frac{{\rm d}\bm{\Psi}(\bm{q},t)}{{\rm d}t} ~,
\end{align}
where $a$ is a scale factor of the universe.
In the Zel'dovich approximation, the displacement field is related to the (Lagrangian) linear density field $\delta_{\rm L}(\bm{q},t)$ through
\begin{equation}
\bm{\nabla}\cdot\bm{\Psi}_{\rm ZA}(\bm{q},t) = - \delta_{\rm L}(\bm{q},t) ~.
\label{eq:psi_to_delta_ZA}
\end{equation}
Recalling that $\delta_{\rm L}$ is expressed as $\delta_{\rm L} = D_{+}(t)\delta_{0} $ with $D_{+}$ and $\delta_{0}$ being respectively the linear growth factor and initial density field, the velocity field in the Zel'dovich approximation is written as
\begin{equation}
\bm{v}(\bm{x}) = aHf(t)\bm{\Psi}_{\rm ZA}(\bm{q},t) ~,
\end{equation}
where $H$ is Hubble parameter and $f(t)$ is a linear growth rate defined by
\begin{equation}
f(t) \equiv \frac{{\rm d}\ln{D_{+}(a)}}{{\rm d}\ln{a}}~.
\end{equation}
Hereafter we omit the subscript $_{\rm ZA}$ and the time-dependence from the argument of functions.

\subsection{Modelling the gravitational redshift contribution}
\label{subsec:modelling_potential}

With the Zel'dovich approximation, the Eulerian redshift-space position $\bm{s}$, given at Eq.~(\ref{eq: s to x}), is mapped into the Lagrangian position $\bm{q}$.
The relativistic correction, $\bm{\epsilon}(\bm{x})$, is also mapped from the Eulerian to Lagrangian space, and we have $\bm{\epsilon}(\bm{x})\simeq \epsilon(\bm{q})\,\hat{\bm{q}}$ at leading order.
Then, using the Poisson equation, one may relate the quantity $\epsilon$ with the linear density field $\delta_{\rm L}$ through Eq.~(\ref{eq:epsilon_phi}). However, in computing the cross-correlation function, we have to relate the redshift-space position of halo/galaxy to the Lagrangian space counterpart. 
Since the halo and galaxy are likely to be formed under the deep potential well via the nonlinear processes, the quantity $\epsilon$ or gravitational potential $\phi$ at the halo/galaxy position would not be simply characterized by the linear density field.
Rather, it would involve a non-perturbative correction, and the quantity $\epsilon$ have to be decomposed into two pieces:
\begin{align}
\epsilon(\bm{x}) &= \epsilon_{\rm L}(\bm{x}) + \epsilon_{\rm NL, X}~, \label{eq: epsilon + NL epsilon}
\end{align}
where the first term $\epsilon_{\rm L}(\bm{x})$ represents the linear-order contribution arising from the gravitational potential of the linear density field, $\phi_{\rm L}(\bm{x})$:
\begin{align}
\epsilon_{\rm L}(\bm{x}) &= - \frac{\phi_{\rm L}(\bm{x})}{aH},\quad
\nabla^{2}\phi_{\rm L}(\bm{x}) = \frac{3\Omega_{\rm m0}H^{2}_{0}}{2a}\delta_{\rm L}(\bm{x}) ~.\label{eq: Poisson eq}
\end{align}
On the other hand, the second term at the right-hand side of Eq.~(\ref{eq: epsilon + NL epsilon}) describes the non-perturbative contribution, arising from the nonlinear density field:
\begin{align}
\epsilon_{\rm NL,X} &= - \frac{\phi_{\rm NL, X}}{aH} ~
\label{eq: non-linear potential}
\end{align}
with $\phi_{\rm NL, X}$ particularly characterizing the non-perturbative part of the potential for the halo/galaxy population ${\rm X}$.
In principle, the halo potential possesses random nature whose statistical properties are determined by the formation process, and thus its amplitude/depth might have some environmental dependences. However, in modelling the cross-correlation function of haloes, which we will later make a detailed comparison with $N$-body simulations, we simply assume that Eq.~(\ref{eq: non-linear potential}) is constant, dependent only on the halo mass and redshift.

One simple prescription to analytically predict $\epsilon_{\rm NL,X}$ is to use the universal halo density profile called Navarro-Frenk-White (NFW) profile~\citep{1996ApJ...462..563N}.
Given the halo mass $M$ and redshift $z$, the potential at halo centre $\overline{\phi}_{\rm NFW}(z,M)$ is analytically computed from the NFW profile.
In Appendix~\ref{appendix: halo potential}, the explicit expressions to compute $\overline{\phi}_{\rm NFW}$ are summarized.
One can thus relate $\overline{\phi}_{\rm NFW}$ to the non-perturbative potential term $\phi_{\rm NL, X}$, i.e., $\phi_{\rm NL, X} = \overline{\phi}_{\rm NFW}(z,M_{\rm X})$.
To be precise, the non-perturbative term is defined by the correction to the linear gravitational potential in Eq.~(\ref{eq: epsilon + NL epsilon}). That is, for a consistent treatment, one needs to subtract the linear contribution from the non-perturbative contribution. However, we have checked and confirmed that the ratio of the linear to non-perturbative potential, $\phi_{\rm L}/\overline{\phi}_{\rm NFW}$, is typically $0.06$, and hence the linear contribution to the non-perturbative potential is negligible. In what follows, we will compute the non-perturbative potential from the NFW profile without subtracting the linear potential contribution.
We will quantitatively check the prescription based on the NFW profile in comparison with $N$-body simulations in Sec.~\ref{sec: results}.

\subsection{Wide-angle redshift-space correlation function with gravitational redshift}

Provided all the ingredients, we are in position to present an analytic model of the redshift-space cross-correlation function with wide-angle and gravitational redshift effects, following \citet{2020MNRAS.491.4162T}.
Let us substitute Eqs.~(\ref{eq: s to x potential}), (\ref{eq: x to q}), and (\ref{eq: epsilon + NL epsilon}) into Eq.~(\ref{eq: s to x}).
We obtain
\begin{align}
s_{i} &= q_{i} +
\Bigl( \delta_{ij}+f \hat{x}_i\hat{x}_{j} \Bigr) \Psi_{j}(\bm{q}) + \epsilon_{\rm L}(\bm{x}) \hat{x}_{i}
 + \epsilon_{\rm NL,X} \hat{x}_{i}
\notag
\\
&\simeq q_{i} + \Psi^{(\rm S)}_{i}(\bm{q}) + \epsilon_{\rm NL,X} \hat{q}_{i}~.
\label{eq: mapping_s_q-space epsilon}
\end{align}
Here, we define the new displacement field $\bm{\Psi}^{(\rm S)}$, including the relativistic contribution:
\begin{align}
\Psi^{(\rm S)}_{i}(\bm{q}) &\equiv 
R_{ij}(\hat{\bm{q}}) \Psi_j(\bm{q}) + \epsilon_{\rm L}(\bm{q}) \hat{q}_{i} ~,
\label{eq: Psi RSD} \\
R_{ij}(\hat{\bm{q}}) &= \delta_{ij} + f \hat{q}_{i}\hat{q}_{j} ~.
\end{align}
The second line of Eq.~(\ref{eq: mapping_s_q-space epsilon}) is valid as long as $\bm{\Psi}$, $\epsilon_{\rm L}$, and $\epsilon_{\rm NL,X}$ are small.
Note that in the distant-observer limit, the unit vector $\hat{\bm{q}}$ in Eq.~(\ref{eq: mapping_s_q-space epsilon}) is replaced with the fixed direction, $\hat{\bm{z}}$.

On the basis of the mapping relation at Eq.~(\ref{eq: mapping_s_q-space epsilon}), the expression of the two-point cross-correlation function is derived, assuming the linear bias relation for all objects to cross correlate.
The detailed derivation is presented in Appendix~\ref{appendix: correlation function}.
The resultant expression for the cross-correlation between different species X and Y becomes~[see Eqs.~(12)--(14) in \citealt{2020MNRAS.491.4162T}]
\begin{equation}
1 + \xi^{(\rm S)}_{\rm XY}(\bm{s}_{1},\bm{s}_{2})
= \frac{\DD(\bm{s}_{1},\bm{s}_{2})}{\RX(\bm{s}_{1}) \RY(\bm{s}_{2})} ~,
\label{eq: def xi XY}
\end{equation}
where the numerator and denominator, $\DD$ and $\RX$ (or $\RY$), are respectively given by
\begin{align}
\DD(\bm{s}_{1},\bm{s}_{2})
&= 
\int{\rm d}^{3}q_{1}
\int{\rm d}^{3}q_{2}
\int\frac{{\rm d}^{3}k_{1}}{(2\pi)^{3}}
\int\frac{{\rm d}^{3}k_{2}}{(2\pi)^{3}}
\notag \\
&\times 
{\rm e}^{{\rm i}\bm{k}_{1}\cdot \left( \bm{s}_{1} - \bm{q}_{1} - \epsilon_{\rm NL, X} \hat{\bm{q}}_{1}\right)
+{\rm i} \bm{k}_{2}\cdot \left( \bm{s}_{2} - \bm{q}_{2} - \epsilon_{\rm NL, Y} \hat{\bm{q}}_{2}\right)}
\notag \\
&\times 
\Biggl<
{\rm e}^{- {\rm i}\bm{k}_{1}\cdot \bm{\Psi}^{(\rm S)}(\bm{q}_{1})
- {\rm i}\bm{k}_{2}\cdot \bm{\Psi}^{(\rm S)}(\bm{q}_{2}) }
\notag \\
&\times 
\left( 1 + b^{\rm L}_{\rm X}\delta_{\rm L}(\bm{q}_{1})\right)
\left( 1 + b^{\rm L}_{\rm Y}\delta_{\rm L}(\bm{q}_{2})\right)
\Biggr> ~,
\label{eq: DD}
\end{align}
\begin{align}
\RX(\bm{s}_{1})	
=
\int{\rm d}^{3}q\; 
&
\int\frac{{\rm d}^{3}k}{(2\pi)^{3}}\;
{\rm e}^{{\rm i}\bm{k}\cdot \left( \bm{s}_{1} - \bm{q} - \epsilon_{\rm NL, X} \hat{\bm{q}}\right)}
\notag \\
&\times 
\Braket{
\left( 1 + b^{\rm L}_{\rm X}\delta_{\rm L}(\bm{q})\right)
{\rm e}^{- {\rm i}\bm{k}\cdot \bm{\Psi}^{(\rm S)}(\bm{q})}
} ~,
\label{eq: RX braket}
\end{align}
where the bracket $\Braket{\cdots}$ stands for the ensemble average, and $b^{\rm L}_{\rm X}$ and $b^{\rm L}_{\rm Y}$ are the Lagrangian linear bias parameters for the population X and Y, respectively.
Note that in Eq.~(\ref{eq: def xi XY}), the mean number density is assumed to be constant over the survey region. Strictly, our analytical model ceases to be adequate if  we consider the non-uniform selection function. Nevertheless, as long as the mean number density does not change drastically, we expect Eq.~(\ref{eq: def xi XY})  to be valid, applicable to the non-uniform case \citep[see discussion in Sec.~3.3.2 of][]{2020MNRAS.491.4162T}.

Setting all the relativistic contributions to zero (i.e, $\epsilon_{\rm NL, X}=\epsilon_{\rm NL, Y}=\epsilon_{\rm L}=0$), the expressions given above exactly coincide with those of \citet{2020MNRAS.491.4162T}. 
The function $R_{\rm X, Y}$ becomes unity if we take the distant-observer limit and ignore the relativistic correction.
Indeed, dropping the higher-order terms of $\mathcal{O}\left( \epsilon^{2}_{\rm NL,X/Y} \right)$, $\mathcal{O}\left( \epsilon_{\rm NL, X/Y} P_{\rm L} \right)$ and $\mathcal{O}\left(P_{\rm L}^2 \right)$, the expressions given above consistently reproduce the linear-order expressions known in the literature, including both the wide-angle and gravitational redshift effects~\citep[e.g.,][]{2009JCAP...11..026M,2014PhRvD..89h3535B}.
In Appendix~\ref{appendix: linearization}, we explicitly demonstrate this and show that our model prescription is a consistent nonlinear extension based on the Zel'dovich approximation.

The cross-correlation function given at Eq.~(\ref{eq: def xi XY}) is still an intricate expression involving the higher-dimensional integrals, which looks difficult to evaluate.
Fortunately, making use of the Gaussianity of the linear density field, part of the integrals can be analytically performed, as shown in \citet{2020MNRAS.491.4162T} (see Appendix~\ref{appendix: correlation function} for derivation). The final form of the expression for the $\RX$ (or $\RY$) becomes
\begin{align}
\RX(\bm{s}_{1})	
= 
\int&\frac{{\rm d}^{3}q}{(2\pi)^{3/2}|\det{A}|^{1/2}}\;
\notag \\
&\times 
{\rm e}^{-\frac{1}{2}\left( s_{1i}-q_{i}- \epsilon_{\rm NL, X} \hat{q}_{i} \right)\left( s_{1j}-q_{j}- \epsilon_{\rm NL, X} \hat{q}_{j} \right)
A^{-1}_{ij}}
\notag \\
&\times 
\left( 1+ b^{\rm L}_{\rm X}\left( s_{1j}-q_{j}- \epsilon_{\rm NL, X} \hat{q}_{j}\right)A^{-1}_{ij}y_{i}\right) ~.
\label{eq: RXRY final}
\end{align}
Here, $A_{ij}$ and $y_{i}$ are defined in Eqs.~(\ref{eq: Aij def}) and (\ref{eq: yi def}), respectively.

To give a simplified expression for $\DD$, we introduce the six-dimensional vectors $\bm{Q} \equiv (\bm{q}_{1}, \bm{q}_{2})$, $\bm{S} \equiv (\bm{s}_{1}, \bm{s}_{2})$, and $\bm{E} \equiv \left( \epsilon_{\rm NL, X} \hat{\bm{q}}_{1}, \epsilon_{\rm NL, Y} \hat{\bm{q}}_{2} \right)$. Then the numerator $\DD$ is expressed as follows:
\begin{align}
\DD & (\bm{s}_{1},\bm{s}_{2})
= 
\int\frac{{\rm d}^{6}Q}{(2\pi)^{3}|\det{\mathcal{A}}|^{1/2}}
{\rm e}^{-\frac{1}{2}(S-Q-E)_{a}(S-Q-E)_{b}\mathcal{A}^{-1}_{ab}} \notag \\
&\times
\Biggl[
1 + b^{\rm L}_{\rm X}b^{\rm L}_{\rm Y}\xi_{\rm L}(q)
-\mathcal{A}^{-1}_{ab}\mathcal{U}_{a}(S-Q-E)_{b}
\notag \\
&
- \Bigl( \mathcal{A}^{-1}_{ab}-\mathcal{A}^{-1}_{ac}\mathcal{A}^{-1}_{bd}(S-Q-E)_{c}(S-Q-E)_{d}
\Bigr)\mathcal{W}_{ab}
\Biggr] ~,
\label{eq: DXDY final}
\end{align}
with the subscripts $a$, $b$, $c$, and $d$ running over 1--6.
The quantities $\xi_{\rm L}$, $\mathcal{A}_{ab}$, $\mathcal{U}_{a}$, and $\mathcal{W}_{ab}$ are defined at Eqs.~(\ref{eq: def xi_L}), (\ref{eq: def mathcal A}), (\ref{eq: def mathcal U}), and (\ref{eq: def mathcal W}), respectively.

Given the linear matter power spectrum $P_{\rm L}$, Eulerian bias parameters $b_{\rm X/Y}$, and non-perturbative part of the potential $\phi_{\rm NL,X/Y}$ (or $\epsilon_{\rm NL,X/Y}$), the prediction of the cross-correlation function is made with Eqs.~(\ref{eq: def xi XY}) through the explicit calculations of Eqs.~(\ref{eq: RXRY final}) and (\ref{eq: DXDY final}). The remaining three- and six-dimensional integrals in the functions $\RX$ (and $\RY$) and $\DD$ are to be evaluated numerically. In the analysis presented below, we adopt the multi-dimensional integration technique of {\tt cuhre} routine in the {\tt CUBA} library to quantitatively compute the cross-correlation function~\citep{2005CoPhC.168...78H}\footnote{\url{http://www.feynarts.de/cuba/}}.

\section{Predictions of dipole cross-correlation function}
\label{sec: model prediction}

In this section, focusing on how the gravitational redshift effect changes the clustering amplitude, we compute Eq.~(\ref{eq: def xi XY}) with (\ref{eq: RXRY final}) and (\ref{eq: DXDY final}), and present the quantitative predictions of dipole cross-correlation function for haloes, which will be later compared with the results of numerical simulations in Sec.~\ref{sec: results}.
Given a set of cosmological parameters, the remaining parameters needed for our model predictions are the redshift, halo bias, and non-perturbative potential, the latter of which is, in this section, computed with the NFW profile for a given halo mass and redshift.
In Table~\ref{tab: population}, we summarize their numerical values for each halo dataset.
Note that the name of the dataset, \data{N}, represents the catalogue of halos having the masses in the range $[N\,m_{\rm DM},\,2N\,m_{\rm DM}]$, with $m_{\rm DM}=1.88\times10^{10}\,h^{-1}M_\odot$ being the mass of dark matter particle.

In Sec.~\ref{sec: effects of phi NL}, we first discuss the impact of the non-perturbative contribution to the gravitational redshift effect introduced in Sec.~\ref{subsec:modelling_potential}.
Then, we investigate the bias and redshift dependence of the dipole cross-correlation in Sec.~\ref{sec: bias and redshift dep}.

Throughout the analysis, the bias parameter of the object $X$ is always taken to be larger than that of the object $Y$ (i.e., $b_{\rm X}>b_{\rm Y}$) so that the separation vector $\bm{s}$ always points from the massive to less massive haloes, as shown in Fig.~\ref{fig: configuration}.

\begin{figure}
\centering
\includegraphics[width=\columnwidth]{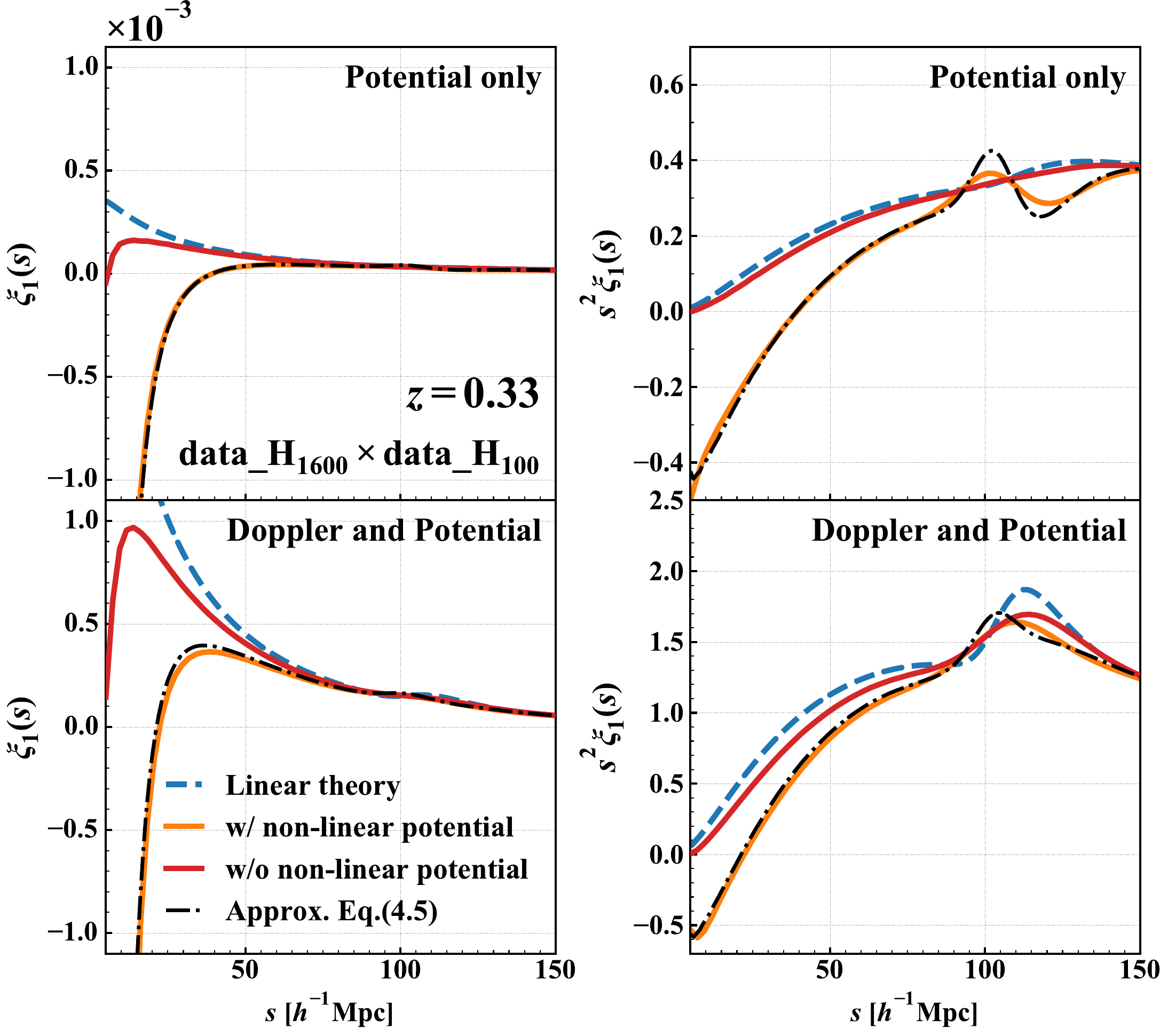}
\caption{Dipole moment of the cross-correlation function between \data{1600} and \data{100} at $z=0.33$.
In left panel, the cross-correlation functions are directly shown, while in right panel, the results are multiplied by the square of separation (i.e., $s^2\xi_1$) in order to make the large-scale behaviors more visible.
{\it Top panel}: only the contribution from the gravitational redshift effect.
{\it Bottom panel}: the contributions from the Doppler and gravitational redshift effects.
We show the predictions based on the linear theory~(blue-dashed), quasi-linear model with the non-linear potential based on NFW profile~(orange), and without the non-linear potential~(red).
The black dot-dashed lines indicate the approximation given at Eq.~(\ref{eq: explanation negative dipole}) (see Sec.~\ref{sec: approximate formula}).
}
\label{fig: xi halo potential}
\end{figure}

\begin{table*}
\scriptsize
\caption{Summary of the dataset: name (\data{N}), mean redshift of \data{1600}, number of haloes, Eulerian linear bias measured by $N$-body simulation, non-linear potential based on NFW profile: $\overline{\phi}_{\rm NFW}$ (Eq.~(\ref{eq: phibar NFW})) and simulations: $\overline{\phi}_{\rm sim}$ (Sec.~\ref{sec: halo potential sim}).}
\begin{tabular}{cccccc}
name
 & redshift
 & number of haloes ($\times 10^{6}$)
 & bias
 & $\overline{\phi}_{\rm NFW}$ ($\times (-10^{-5})$)
 & $\overline{\phi}_{\rm sim}$ ($\times (-10^{-5})$)
\\
\hline\hline
 & 0.33 & $5.4$ & 1.08 & 0.285 & 0.620 \\
 & 0.9 & $6.2$ & 1.60 & 0.325 & 0.606 \\
\data{100} & 1.1 & $7.0$ & 1.68 & 0.347 & 0.604 \\
 & 1.3 & $7.4$ & 2.00 & 0.373 & 0.600 \\
 & 1.5 & $7.3$ & 2.19 & 0.404 & 0.595 \\
\hline
 & 0.33 & $3.4$ & 1.22 & 0.439 & 0.711 \\
 & 0.9 & $0.36$ & 1.98 & 0.509 & 0.727 \\
\data{200} & 1.1 & $0.39$ & 2.03 & 0.547 & 0.728 \\
 & 1.3 & $0.40$ & 2.23 & 0.591 & 0.732 \\
 & 1.5 & $0.38$ & 2.55 & 0.643 & 0.735 \\
\hline
 & 0.33 & $1.9$ & 1.42 & 0.678 & 0.848 \\
 & 0.9 & $0.18$ & 2.28 & 0.800 & 0.898 \\
\data{400} & 1.1 & $0.19$ & 2.36 & 0.863 & 0.915 \\
 & 1.3 & $0.18$ & 2.87 & 0.938 & 0.941 \\
 & 1.5 & $0.16$ & 3.05 & 1.03 & 0.970 \\
\hline
 & 0.33 & $0.96$ & 1.69 & 1.05 & 1.06 \\
 & 0.9 & $0.080$ & 2.63 & 1.26 & 1.15 \\
\data{800} & 1.1 & $0.078$ & 2.80 & 1.37 & 1.21 \\
 & 1.3 & $0.069$ & 3.54 & 1.49 & 1.24 \\
 & 1.5 & $0.057$ & 3.84 & 1.64 & 1.29 \\
\hline
 & 0.33 & $0.44$ & 2.07 & 1.63 & 1.37 \\
 & 0.9 & $0.030$ & 3.49 & 1.98 & 1.56 \\
\data{1600} & 1.1 & $0.027$ & 3.70 & 2.16 & 1.63 \\
 & 1.3 & $0.022$ & 4.17 & 2.37 & 1.67 \\
 & 1.5 & $0.016$ & 4.83 & 2.62 & 1.76 \\
\end{tabular}
\label{tab: population}
\end{table*}

\subsection{Impact of non-perturbative halo potential}
\label{sec: effects of phi NL}

To see the significance of non-perturbative correction to the gravitational redshift effect, the predictions of dipole cross-correlation are presented, both including and ignoring the contribution $\epsilon_{\rm NL,X}$, which we will further compare with linear theory predictions.
In Appendix~\ref{appendix: linear theory}, as an extension of~\citet{2004ApJ...614...51S,2008MNRAS.389..292P}, we derive and summarize the linear theory expressions of the cross-correlation function, and the new contributions from gravitational redshift effect are particularly presented on top of the wide-angle formulas in \citet{2020MNRAS.491.4162T} (see their Appendix D).

In Fig.~\ref{fig: xi halo potential}, the predicted dipoles are shown at $z=0.33$ for the cross-correlation between the most and least massive haloes (\data{1600} and \data{100}).
In right panel, to clarify the large-scale behaviors, the results are multiplied by the square of separation.
The upper panels plot the results taking only the gravitational redshift effect into account as a source of RSD\footnote{In our quasi-linear model, the predictions ignoring the Doppler effect can be made by setting the linear growth rate $f$ to zero.}, while in lower panels, both Doppler and gravitational redshift effects are considered.

On large scales, the quasi-linear predictions are supposed to converge to the linear theory predictions (blue dashed).
This is indeed the case for predictions ignoring the non-perturbative potential (red solid).
On the other hand, in the presence of non-perturbative halo potential (orange solid), as shown in right panel, the actual convergence is very slow.
Though the quantitative differences between predictions are small, the baryon acoustic features around $s\approx 110\; h^{-1}{\rm Mpc}$ exhibit a notable difference.
In particular, in the cases only including the gravitational redshift effect (upper panel), peak and trough structures are more visible in the quasi-linear prediction, rather than in the linear theory prediction.
As we will explain later in Sec.~\ref{sec: approximate formula}, this is ascribed to the modulation of the cross-correlation function due to the non-perturbative halo potential, with which the mixing of the monopole moment appears manifest (see Eq.~(\ref{eq: explanation negative dipole})).

Turning to the dipole moments at small scales (left panel in Fig.~\ref{fig: xi halo potential}), $s\lesssim 50\; h^{-1}{\rm Mpc}$, the predicted dipole including the non-perturbative contribution $\epsilon_{\rm NL,X}$ starts to significantly deviate from others, and its amplitude eventually turns from positive to negative around $s\simeq 40\,h^{-1}{\rm Mpc}$.
This trend can be also seen even if turning on the Doppler effect, shown in the bottom panel of Fig.~\ref{fig: xi halo potential}, although the amplitude of the dipole cross-correlation is, overall, lifted up due to the Doppler effect, and the sign flip of the dipole appears at a smaller scale.

The sign flip of the dipole amplitude seen in the quasi-linear predictions indeed matches those found in the numerical simulations \citep{2019MNRAS.483.2671B}, and thus implies that the non-perturbative contribution indeed plays an important role. While a more quantitative aspect of the analytical predictions will be discussed in detail in Sec.~\ref{sec: results}, we shall also give a simple explanation for the sign flip of the dipole in Sec.~\ref{sec: approximate formula}.

\subsection{Bias and redshift dependencies}
\label{sec: bias and redshift dep}

\begin{figure}
\centering
\includegraphics[width=\columnwidth]{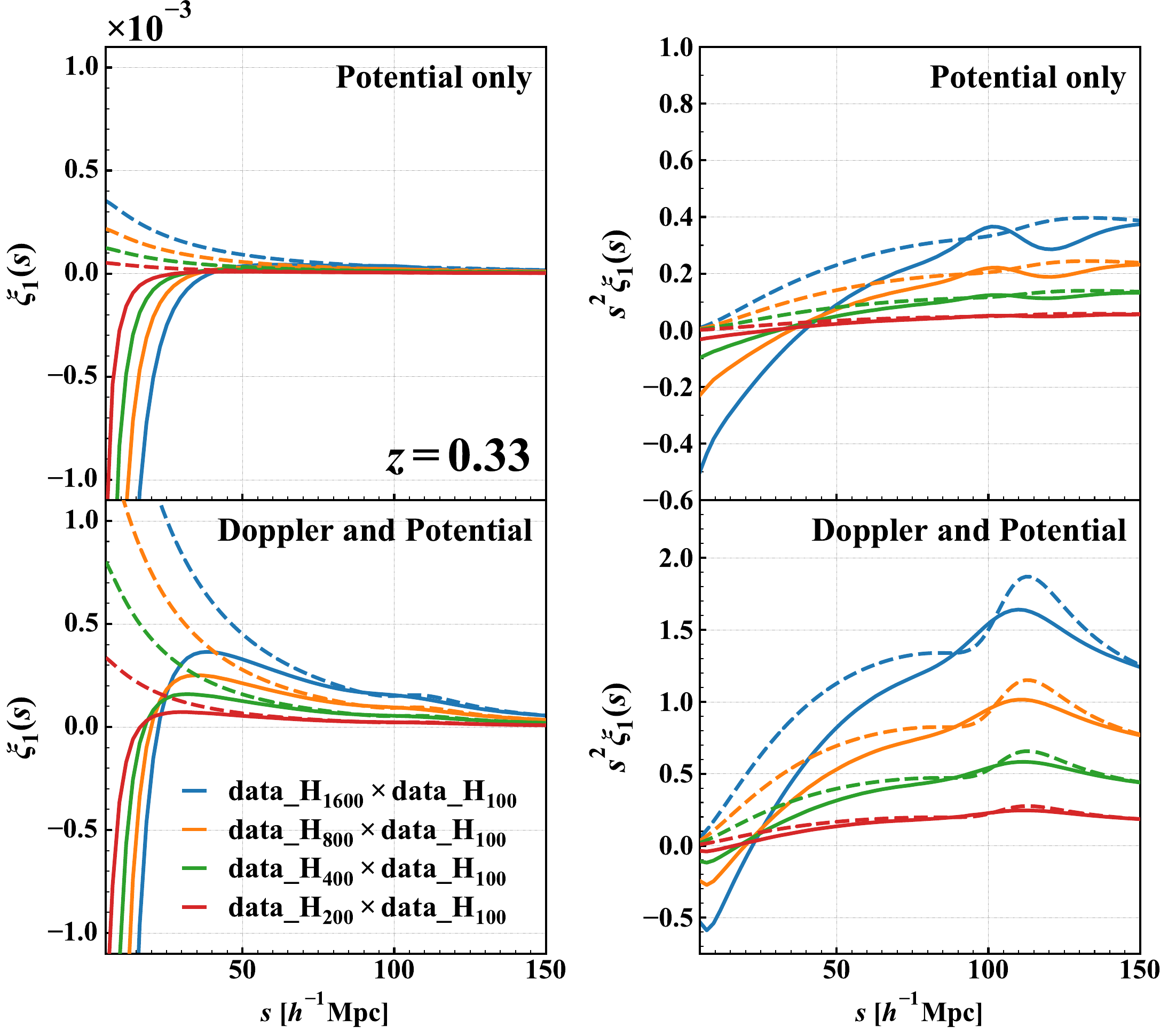}
\caption{Bias dependence of the predicted dipole cross-correlation. While left panels directly plot the results of dipole moments, right panels show the predictions multiplying by the square of separation. Fixing the redshift to $z=0.33$, we take the cross-correlation between haloes of \data{100} and other massive haloes. The upper panels plot the results including only the gravitational redshift effect as a source of RSD, while lower panels show the predictions including both the Doppler and gravitational redshift effects. Solid and dashed lines represent the quasi-linear and linear theory predictions, respectively.}
\label{fig: bias_dependence}
\end{figure}

\begin{figure}
\centering
\includegraphics[width=\columnwidth]{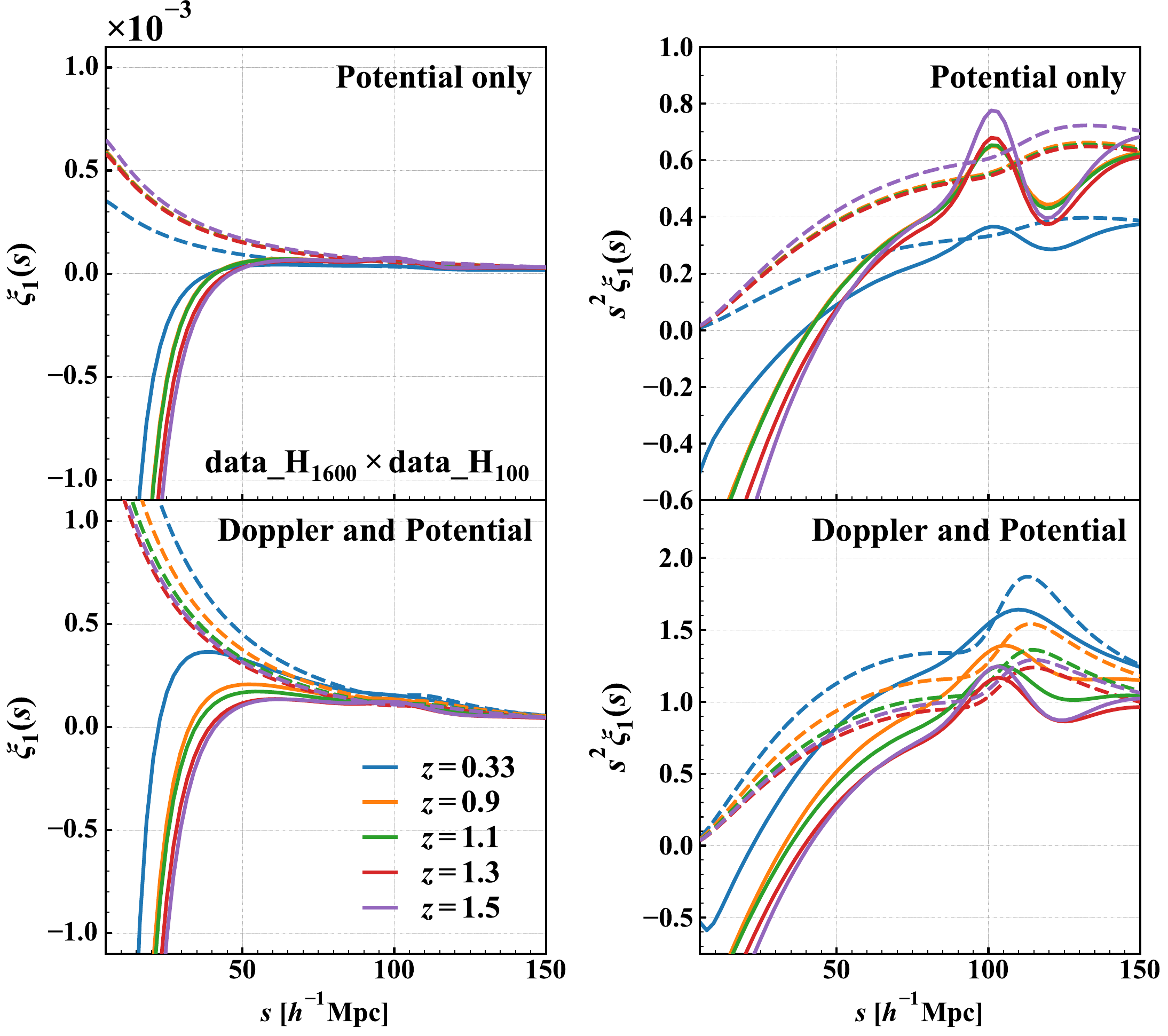}
\caption{
Same as Fig.~\ref{fig: bias_dependence} but for redshift dependence of the predicted dipole cross-correlation. Varying the redshifts from $z=0.33$ to $1.5$, the dipole cross-correlation function between \data{1600} and \data{100} are shown.
}
\label{fig: redshift_dependence}
\end{figure}

Turn next to focus on the bias and redshift dependencies of the dipole cross-correlation function.
In Fig.~\ref{fig: bias_dependence}, fixing the redshift to $z=0.33$, the dipole cross-correlations between haloes of \data{100} and other massive populations with a larger value of halo bias are plotted.
In Fig.~\ref{fig: redshift_dependence}, varying the redshifts from $z=0.33$ to $1.5$, the predicted dipoles are shown for a specific cross-correlation between \data{1600} and \data{100}.
In both cases, the upper and lower panels respectively plot the cases including only the gravitational redshift and both the gravitational redshift and Doppler effects.
Note that in all quasi-linear predictions, the non-perturbative contribution to the gravitational redshift effect is included.

In Fig.~\ref{fig: bias_dependence}, the predicted dipoles systematically change with the difference between the biases for cross-correlating haloes.
Through the monotonic mass dependence of the non-perturbative halo potential (see Table \ref{tab: population}), the dipole arising from the gravitational redshift effect shows a sharper drop and exhibits the sign flip at a larger scale as increasing the bias difference (upper panel).
Including the Doppler effect, the overall amplitude gets enhanced due to the linear dependence of the Doppler-induced dipole, but the sign flip still remains manifest.
This is because the contribution from the gravitational redshift effect also increases with bias difference.

On the other hand, in Fig.~\ref{fig: redshift_dependence} showing the redshift dependence, we do not clearly see a systematic trend when including the Doppler and gravitational redshift effects, especially at large scales.
This non-monotonic behavior comes from the two competitive effects deduced from the linear theory.
The gravitational redshift contribution in the linear theory is expressed, at the leading order of $O(s/d)$, as~[Eqs.~(\ref{eq: wa expansion app}) and (\ref{eq: xi 1 0})]
\begin{equation}
\xi^{(\rm S)}_{1,{\rm lin}}(s,d) \simeq (b_{\rm X} - b_{\rm Y}) \eta^{3}_{1}(s) ~, \label{eq: grav lin dipole}
\end{equation}
where the function $\eta^{3}_{1}(s)$ is defined in Eq.~(\ref{eq: def eta ell m}), and is proportional to the matter power spectrum (cf. Eq.~(D29) of \citet{2020MNRAS.491.4162T} for the standard Doppler contribution at leading order). 
As we see from Table~\ref{tab: population}, the bias for a mass-selected halo sample gets increased with redshift, and thus the difference of the bias for cross-correlation, $b_{\rm X}-b_{\rm Y}$, tends to be large.
However, from Eq.~(\ref{eq: grav lin dipole}), the amplitude of correlation function linearly depend on the matter power spectrum, and is thus proportional to the square of linear growth factor, which gets decreased with redshift. 
These produce the non-monotonic behaviours in the dipole, and to some extent, they also change the amplitudes of the negative dipole at small scales.

\subsection{Approximate expression for the quasi-linear model}
\label{sec: approximate formula}

Here, we discuss the role of the non-perturbative halo potential contribution in more detail, and exploit an approximate expression that quantitatively describes the sign flip of the dipole cross-correlation function.

First recall that the non-perturbative contribution to the gravitational redshift, $\epsilon_{\rm NL,X}=-\phi_{\rm NL,X}/(aH)$, induces a constant offset along the line of sight in the redshift-space positions for each halo population.
This suggests that the cross-correlation function including the non-perturbative contributions, $\xi^{(\rm S)}_{\rm XY}$, may be described by the (nonlinear) cross-correlation function without the halo potential, which we denote by $\widetilde{\xi}^{({\rm S})}_{\rm XY}$. That is, the function $\xi^{(\rm S)}_{\rm XY}$ is given by $\widetilde{\xi}^{({\rm S})}_{\rm XY}$, but the 
redshift-space positions in its argument are shifted away by the gravitational redshift effect from the halo potential, i.e., $\bm{s}_{1,2}\to\bm{s}_{1,2}+\phi_{\rm NL,X/Y}/(aH)\hat{\bm{s}}_{1,2}$\footnote{Seemingly, the sign of the shift by the halo potential looks opposite, but recalling the fact that the positions $\bm{s}_{1,2}$ are those in the cross-correlation $\xi^{(\rm S)}_{\rm XY}$ including the halo potential, this is correct.}:
\begin{align}
\xi^{(\rm S)}_{\rm XY}(\bm{s}_{1},\bm{s}_{2})
&= \widetilde{\xi}^{(\rm S)}_{\rm XY} \left( \bm{s}_{1}+\frac{\phi_{\rm NL,X}}{aH}\hat{\bm{s}}_{1},\, \bm{s}_{2}+\frac{\phi_{\rm NL,X}}{aH}\hat{\bm{s}}_{2} \right) 
~,
\label{eq:xi_model}
\end{align}
The proposition given above is in fact exact if replacing the unit vectors $\hat{\bm{s}}_{1,2}$ in the argument with the fixed line-of-sight direction $\hat{\bm{d}}$.
Expanding Eq.~(\ref{eq:xi_model}) in powers of the halo potential $\phi_{\rm NL,X/Y}$, we obtain
\begin{align}
\xi^{(\rm S)}_{\rm XY}(\bm{s}_{1},\bm{s}_{2})&\simeq 
\left\{ 1+\frac{\phi_{\rm NL,X}}{aH} \left(\hat{\bm{s}}_{1}\cdot\nabla_{s_1}\right) 
+\frac{\phi_{\rm NL,Y}}{aH} \left( \hat{\bm{s}}_{2}\cdot\nabla_{s_2} \right) \right\}
\nonumber
\\
&\times \widetilde{\xi}^{(\rm S)}_{\rm XY}(\bm{s}_{1},\bm{s}_{2})+ \mathcal{O}(\phi_{\rm NL,X/Y}^2) 
\label{eq:xi_expansion_potential}
\end{align}
Writing the redshift-space positions $\bm{s}_1$ and $\bm{s}_2$ with the separation vector $\bm{s}$ and the mid-point line-of-sight vector $\bm{d}=(\bm{s}_1+\bm{s}_2)/2$, Eq.~(\ref{eq:xi_expansion_potential}) leads to
\begin{align}
\xi^{(\rm S)}_{\rm XY}(s,d,\mu)
&\simeq
\left\{ 1- \frac{\Delta \phi_{\rm NL}}{aH} \left( \hat{\bm d}\cdot \nabla_s \right) \right\}\,
\widetilde{\xi}^{(\rm S)}_{\rm XY}(s,d,\mu)
\nonumber
\\
&=
 \left\{ 1-\frac{\Delta \phi_{\rm NL}}{aH} 
\left( \mu \frac{\partial }{\partial s} + \frac{1-\mu^{2}}{s} \frac{\partial}{\partial \mu}\right)
\right\} \widetilde{\xi}^{(\rm S)}_{\rm XY}(s,d,\mu) ~,
\label{eq: xi phi NL leading order}
\end{align} 
where the variable $\mu$ is the directional cosine given by $\mu=\hat{\bm{s}}\cdot\hat{\bm{d}}$, and we define $\Delta\phi_{\rm NL}\equiv \phi_{\rm NL, X}-\phi_{\rm NL, Y}$. In the first line of Eq.~(\ref{eq: xi phi NL leading order}), we assumed that the corrections from the wide-angle effects are small, and dropped the contributions of $\mathcal{O}((s/d)^2)$ and $\mathcal{O}((s/d)\Delta\phi_{\rm NL})$.

Applying the multipole expansion of Eq.~(\ref{eq: multipole expansion}) to Eq.~(\ref{eq: xi phi NL leading order}), the dipole moment ($\ell=1$) of the cross-correlation function becomes
\begin{align}
\xi^{(\rm S)}_{\rm XY, 1}(s,d) \simeq
\widetilde{\xi}^{(\rm S)}_{\rm XY, 1}(s,d)
- \frac{\Delta \phi_{\rm NL}}{aH}\,\frac{\partial \widetilde{\xi}^{(\rm S)}_{\rm XY, 0}(s,d)}{\partial s}~. 
\label{eq: explanation negative dipole}
\end{align}
To be strict, the term proportional to $\Delta\phi_{\rm NL}$ includes also the contributions coming from the quadrupole moment ($\ell=2$), but we have checked and confirmed them to be ignorable.
Note that a similar approximation has been applied in the case of the void-galaxy correlation function with the gravitational redshift effect~\citep{2018PhRvD..98d3527N}.

Eq.~(\ref{eq: explanation negative dipole}) describes how the non-perturbative halo potential modulates the dipole cross-correlation induced by the wide-angle RSD and gravitational redshift from the linear-order potential.
In our setup with $b_{\rm X}>b_{\rm Y}$, as we have seen in Fig.~\ref{fig: xi halo potential} (see red solid or blue dashed lines), the first term at the right-hand side of Eq.~(\ref{eq: explanation negative dipole}) gives a positive contribution.
On the other hand, at the scales of our interest, the monopole cross-correlation monotonically decreases with the separation, and we have $\partial \widetilde{\xi}^{(\rm S)}_{\rm XY, 0}/\partial s<0$.
Also, taking the halo population $X$ to be more massive than the halo subsample $Y$, we usually have $\Delta\phi_{\rm NL}<0$.
That is, the second term at right-hand side of Eq.~(\ref{eq: explanation negative dipole}) gives a negative contribution.

At large scales, the first term approaches linear theory prediction, keeping the amplitude of $\xi^{(\rm S)}_{\rm XY, 1}$ positive, but with a small modulation by the second term.
Although the first term is the leading contribution to dipole, the second term, arising from the monopole contribution, can still give a small imprint on the large-scale behaviors, making the baryon acoustic feature prominent. This is particularly the case when we ignore the standard Doppler effect, as shown in the right panel of Fig.~\ref{fig: xi halo potential}.
On the other hand, at small scales, the first term ceases to be the leading order.
Instead, the second term gets large, and can dominate the first term, leading to the sign flip of the dipole cross-correlation.

In Fig.~\ref{fig: xi halo potential}, the prediction based on Eq.~(\ref{eq: explanation negative dipole}) is plotted, depicted as black dot-dashed lines.
Despite the fact that Eq.~(\ref{eq: explanation negative dipole}) was derived under several approximations and assumptions, the predicted dipoles remarkably agree well with full quasi-linear predictions including the non-perturbative halo potential. Here, in evaluating Eq.~(\ref{eq: explanation negative dipole}), the dipole and monopole cross-correlation functions in Eq.~(\ref{eq: explanation negative dipole}) are computed with the quasi-linear model and linear theory, respectively.
Note that the results almost remain unchanged even if we use the quasi-linear model to compute the monopole moment. The approximation given at Eq.~(\ref{eq: explanation negative dipole}) provide a way to separately treat the contribution from the non-perturbative halo potential, and may be used for a practical application to observations.

Finally, based on Eq.~(\ref{eq: explanation negative dipole}), the characteristic scale $s_{\rm cross}$ where the sign flip of the dipole cross-correlation happens (i.e., $\xi_{\rm XY,1}^{\rm(S)}(s_{\rm cross})=0$) can be estimated as
\begin{align}
s_{\rm cross} &\simeq \frac{\widetilde{\xi}^{(\rm S)}_{\rm XY, 0}}{\widetilde{\xi}^{(\rm S)}_{\rm XY, 1}} \frac{|\Delta \phi_{\rm NL}|}{aH}
\notag 
\\
&= 30\,\,
\frac{\tilde{\xi}^{(\rm S)}_{\rm XY, 0}/\widetilde{\xi}^{(\rm S)}_{\rm XY, 1}}{10^{2}}
\frac{|\Delta \phi_{\rm NL}|}{10^{-5}} 
\frac{H_{0}(1+z)}{H(z)}
\; h^{-1}{\rm Mpc} ,
\label{eq:scale_sign_flip}
\end{align}
where we used the approximation, 
$\partial \widetilde{\xi}^{(\rm S)}_{\rm XY, 0}/\partial s \simeq - \widetilde{\xi}^{(\rm S)}_{\rm XY, 0}/s$, and adopt a typical value of the monopole-to-dipole ratio, $\widetilde{\xi}^{(\rm S)}_{\rm XY, 0}/\widetilde{\xi}^{(\rm S)}_{\rm XY, 1}$, around $s\gtrsim 30\; h^{-1}{\rm Mpc}$.
Eq.~(\ref{eq:scale_sign_flip}) qualitatively describes how the zero-crossing point changes with the bias, redshift and halo potential.
In particular, the dependence of the bias appears through the monopole-to-dipole ratio, and using the linear theory, the zero-crossing point scales as $s_{\rm cross}\,\propto\,b_{\rm X}b_{\rm Y}/(b_{\rm X}-b_{\rm Y}) \left| \Delta \phi_{\rm NL} \right|$, for a given  redshift.
We have checked that the estimated values of $s_{\rm cross}$ from Eq.~(\ref{eq:scale_sign_flip}) reasonably explain those of the quasi-linear predictions seen in Figs.~\ref{fig: bias_dependence} and \ref{fig: redshift_dependence}.

\section{Comparison with simulations}
\label{sec: results}

We are in a position to make a detailed comparison between analytical predictions and numerical simulations at various redshifts and for different halo populations.

The simulation used for the present work is described in \citet{2019MNRAS.483.2671B}. At low redshift ($z < 0.5$) we use the halo catalogue from the full-sky light-cone. The analysis has been extended to a larger redshift range (up to $z = 2$) by producing a halo catalogue from a narrow light-cone (2500 deg$^2$ aperture). The light-cone is tilted with respect to the simulation axis to minimize replication effect. The main difference between the halo catalogues used for the present work and the one used in \citet{2019MNRAS.483.2671B} is that the interpolation from the gravity mesh uses a Triangular Shaped Cloud interpolation (while it used a Cloud In Cell interpolation in an earlier version, but we have verified that it did not change the final results).

As we briefly mentioned in Sec.~\ref{sec: model prediction}, the halo catalogue we use is divided into six subsamples with logarithmically equal-mass bins, summarized in Table~\ref{tab: population}.
Note again that the name of the dataset, \data{N}, implies the halo catalogue having the masses in the range $[N\,m_{\rm DM},\,2N\,m_{\rm DM}]$.
Here, the (Eulerian) halo bias is estimated from the square root of the ratio of the monopole halo to matter auto-correlation functions, and we checked that the bias derivatives are consistent with those predicted from the halo model of~\citet{1999MNRAS.308..119S}.
In addition to the predicted values of non-perturbative contribution to the halo potential, Table~\ref{tab: population} also lists the measured values of the potential from $N$-body simulations, which we will describe in more detail below.

\subsection{Non-perturbative halo potential}
\label{sec: halo potential sim}

The non-perturbative contribution to the gravitational redshift effect, coming from the halo potential, plays an important role in our analytical model prediction at small scales.
While we model it using the NFW profile, the validity of this prescription needs to be quantitatively checked with $N$-body simulations.

After identifying the haloes in a snapshot at the redshift of interest, we measure the gravitational potential at their centre of mass (denoted by $\phi_{\rm halo}$) by interpolating the potential from the AMR grid. For a single halo, we identify its composite particles and trace them back to their initial Lagrangian positions (their position in the initial snapshot). We measure the initial gravitational potential of these particles at the centre of mass of these particles in the initial snapshot. Then, multiplying by $D_{+}(t)/a(t)$, we obtain the linear halo potential at the time of interest, denoted by $\phi_{\rm L}$.
On the basis of this measurement, we obtain $\phi_{\rm halo}$ and $\phi_{\rm L}$ for many haloes by using the snapshots of a smaller $N$-body simulation (512$^3$ particles) with roughly the same resolution as $\Lambda$CDM RayGalGroupSims.
Note that, using higher-resolution simulations with $2^{3}$ times larger volume and number of particles, we confirm that the results remain unchanged.

\begin{figure}
\centering
\includegraphics[width=0.9\columnwidth]{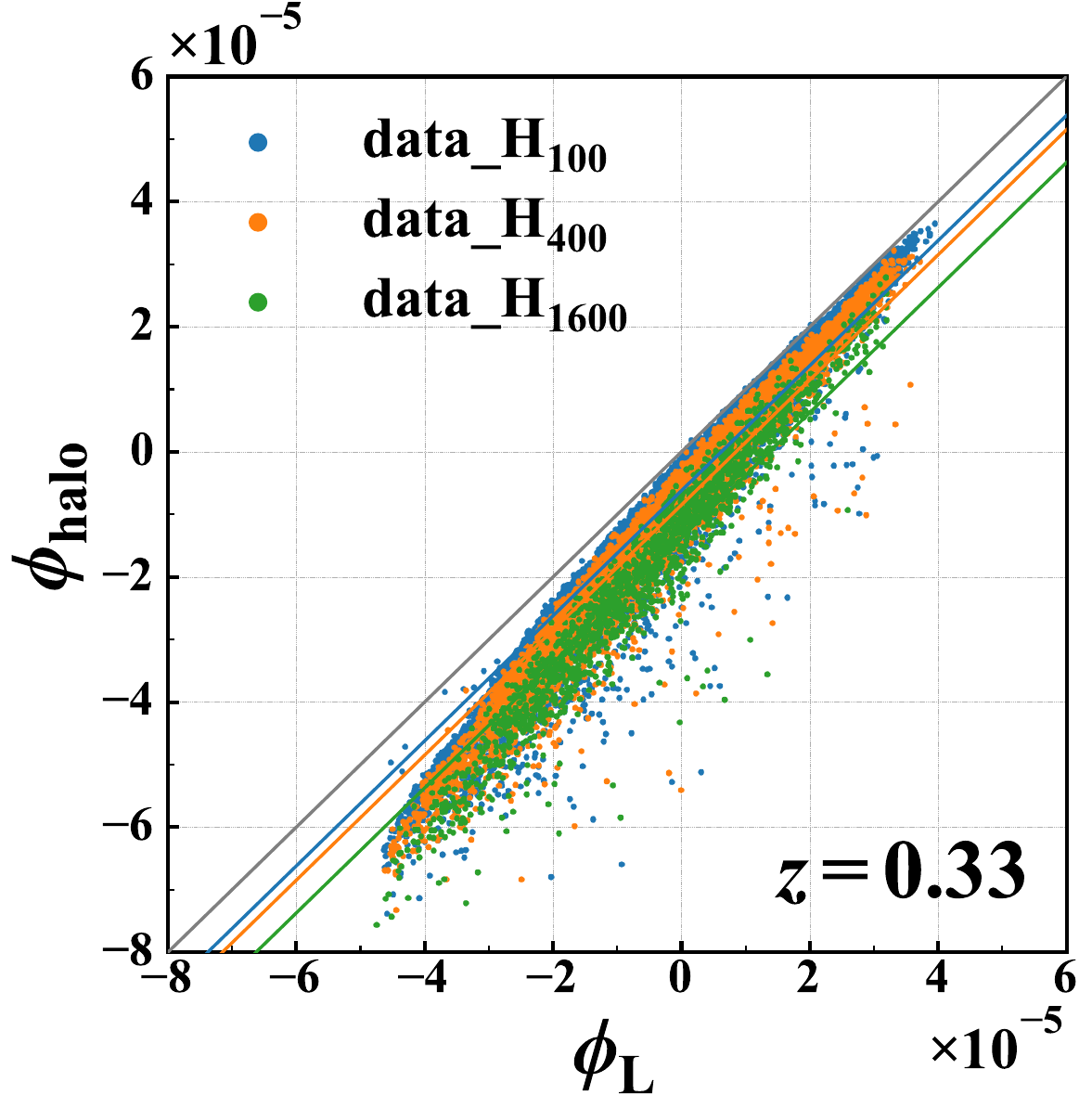}
\caption{Gravitational potential at the centre of mass of haloes $\phi_{\rm halo}$ vs linear potential $\phi_{\rm L}$, estimated from $N$-body simulations for the halo population \data{100}, \data{400}, and \data{1600}.
We show the linear fitting $\phi_{\rm halo} = \phi_{\rm L} + \overline{\phi}_{\rm sim}$.
Grey line indicates $\phi_{\rm halo} = \phi_{\rm L}$.}
\label{fig: phi NL vs L}
\end{figure}

Fig.~\ref{fig: phi NL vs L} shows the measured halo potential, $\phi_{\rm halo}$, against the Lagrangian counterpart, $\phi_{\rm L}$, with the colours indicating the different halo samples. The points represent the measured results from each halo, and we here present the results at $z=0.33$.
Apart from the scatter, the overall trend averaged over each subsample indicates that there is a small offset from the linear relation, $\phi_{\rm halo}=\phi_{\rm L}$, and the offset becomes large as increasing the halo masses. This offset, $\phi_{\rm sim} \equiv \phi_{\rm halo} - \phi_{\rm L}$, can be regarded as the non-perturbative contribution to the gravitational potential. Averaging over all haloes in each dataset of haloes, we obtain $\overline{\phi}_{\rm sim}$, summarized in Table~\ref{tab: population}.

A closer look at Fig.~\ref{fig: phi NL vs L} reveals that the offset of $\phi_{\rm halo}$ from the linear relation $\phi_{\rm halo}=\phi_{\rm L}$ becomes systematically large as decreasing $\phi_{\rm L}$, and rather implies the relation, $\phi_{\rm halo}=\alpha \phi_{\rm L} + \beta$, with $\alpha \gtrsim 1$.
Taking $\alpha$ and $\beta$ as free parameters, we have examined the fitting to this linear relation, and at all redshifts we considered, we obtain $\alpha \approx 1.2$, which is independent of the halo populations.
While this could seemingly give an impact on the prediction of the dipole moment, we checked and confirmed that the predictions of the dipole cross-correlation do not change drastically.
Rather, the additive term, $\beta$ or $\overline{\phi}_{\rm sim}$, plays an important role to predict the dipole moment.
Hence, in what follows, we use the averaged value of $\phi_{\rm sim}$ estimated based on the linear relation $\phi_{\rm halo} = \phi_{\rm L}+\phi_{\rm sim}$.

\begin{figure}
\centering
\includegraphics[width=\columnwidth]{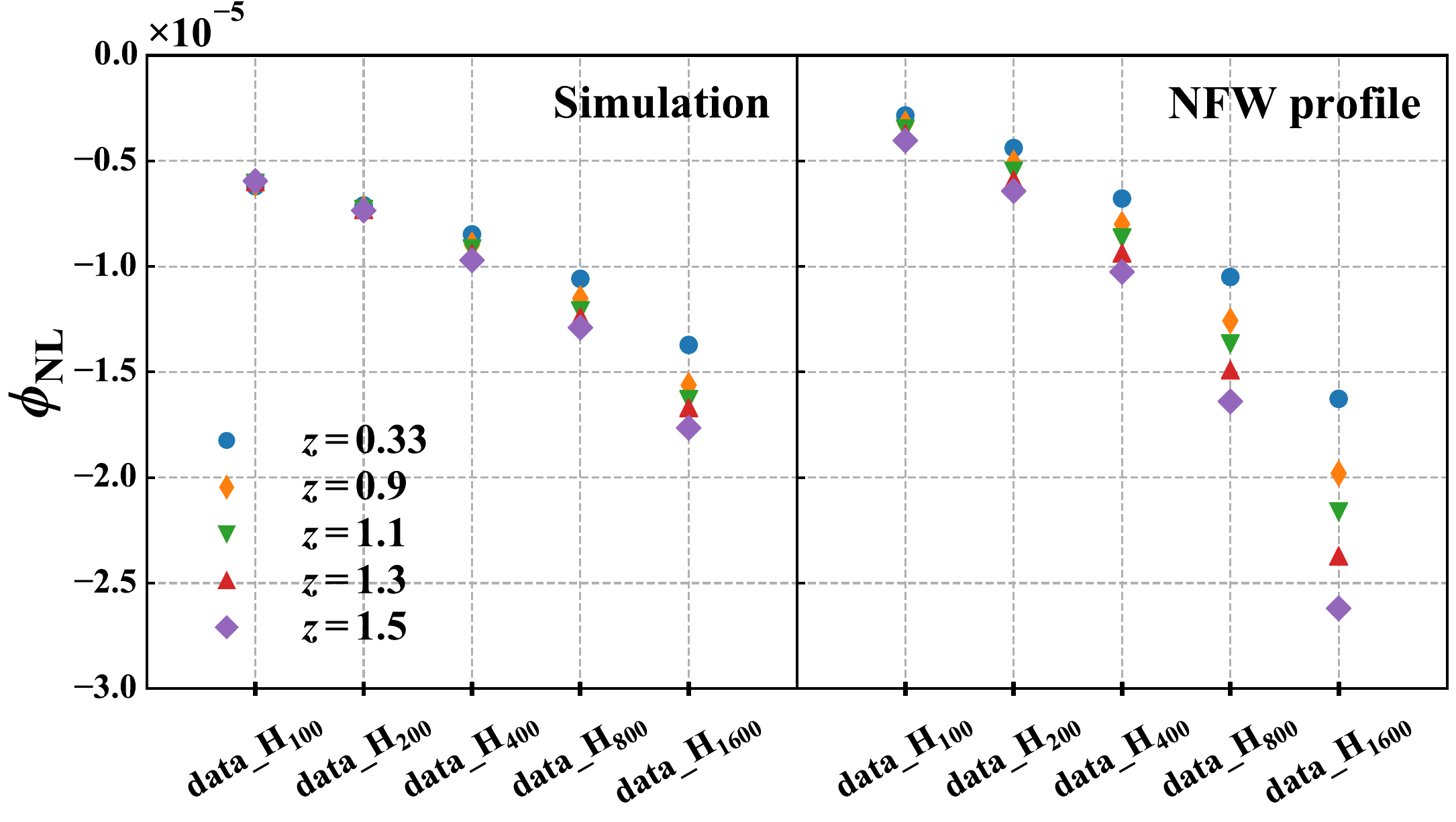}
\caption{
Predictions of the non-linear potential based on $N$-body simulation ({\it left}) and NFW profile ({\it right}).}
\label{fig: halo potential}
\end{figure}

In the left panel of Fig.~\ref{fig: halo potential}, the estimated values of the non-perturbative potential contribution $\overline{\phi}_{\rm sim}$ are shown at various redshifts, and are plotted against the halo subsamples sorted by their mean halo mass.
In the right panel, the analytical model predictions based on the NFW profile are shown.
Qualitatively, both cases show the same trend in the halo mass and redshift dependences.
Quantitatively, however, the model prediction tends to have a shallower potential for less massive halo populations.
Also, compared to the measured results, the prediction exhibits a stronger redshift dependence, and the model eventually predicts a rather deeper potential for massive halo subsamples at higher redshifts.

There are several plausible reasons for the quantitative differences, which are summarized as follows.
First, the model prediction based on the NFW profile ignores the contributions coming from the external fields and/or the influence of nearby haloes.
In simulations, merging of haloes frequently happens and the clustering of haloes themselves develops as decreasing redshifts.
Thus, the evolved halo potential tends to be affected by the surrounding environment in general, and the measured potential can get deeper than that of the model prediction, especially for small haloes.
Second, the estimated value of $\overline{\phi}_{\rm sim}$ is obtained by looking at the potential at the centre-of-mass position for each halo.
However, unlike the NFW profile, the position of the deepest potential generally differs from the centre of mass for aspherical haloes in simulations.
This may result in the underestimation of the non-perturbative contribution, $\phi_{\rm NL}$, especially giving an impact on the massive halo populations.
The third possibility may come from the halo mass definition in $N$-body simulations.
It is known that the mass defined by the Friends-of-Friends (FoF) algorithm does not precisely coincide with that of haloes identified with the Spherical Overdensity finder~(see e.g., \citealt{2009ApJ...692..217L,2011ApJS..195....4M}).
Thus, the halo mass in simulations may differ from that of the spherically symmetric NFW halo, and this can lead to the bias in the potential-halo mass relation, especially for massive haloes. 
Note that, in Appendix \ref{appendix: b = 0.1}, we check the dependence of the halo definition on the non-linear potential by using the different halo definition (i.e. smaller linking length $b = 0.1$). The smaller linking length would lead to a slightly better agreement with the predictions based on spherically symmetric haloes.

For these reasons above, we will below present the quasi-linear predictions, with the non-perturbative halo potential estimated both from simulations and analytical model.

\subsection{Dipole cross-correlation at large scales}

\begin{figure*}
\centering
\includegraphics[width=0.8\textwidth]{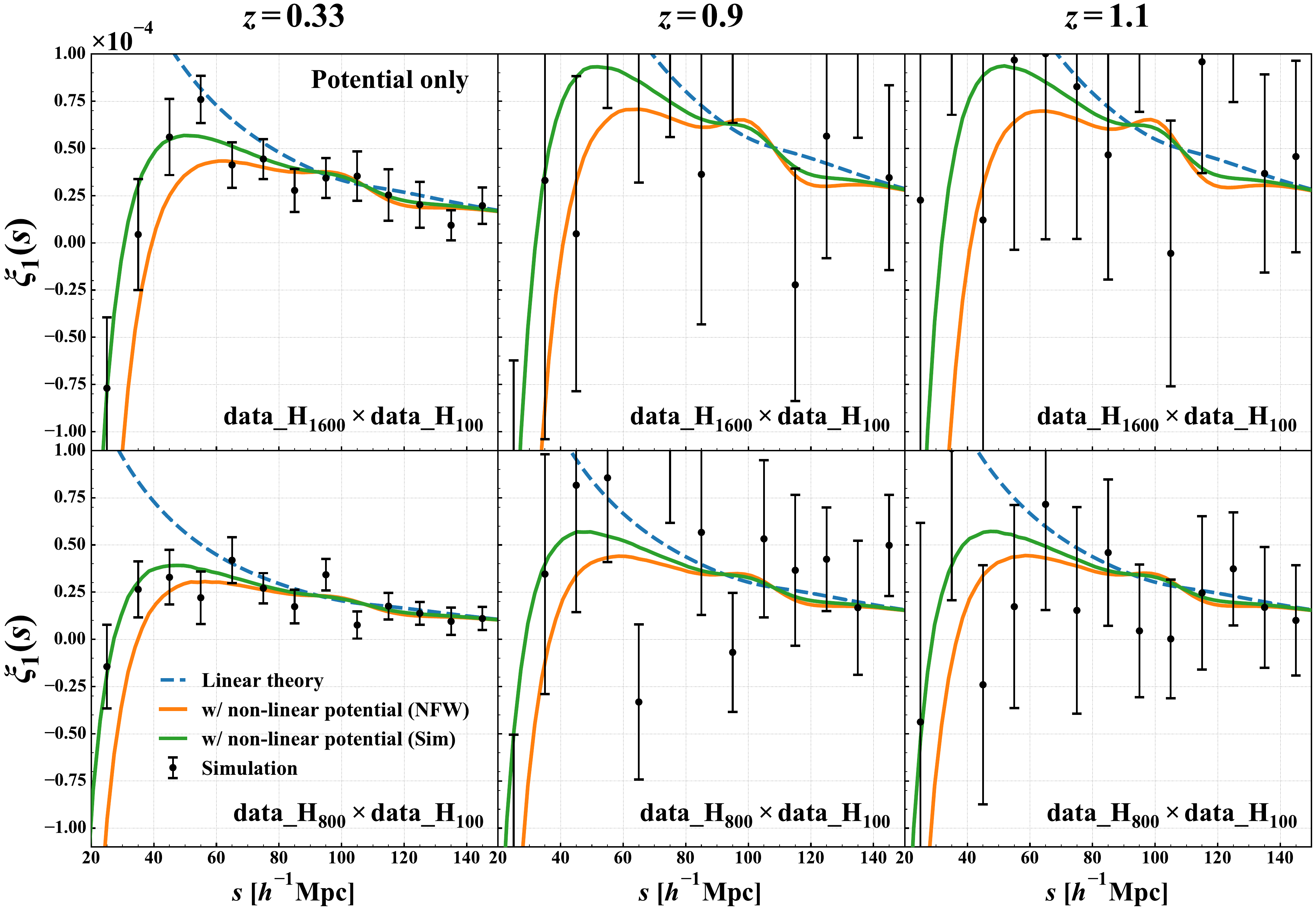}
\caption{The gravitational redshift contribution to the dipole cross-correlation function on large scales at $z=0.33$, $0.9$, and $1.1$ (from {\it left} to {\it right}).
Top and bottom panels show the different halo populations, \data{1600}$\times$\data{100} and \data{800}$\times$\data{100}, respectively.
Among these figures, we show the predictions based on the linear theory~(blue-dashed), quasi-linear predictions with the non-linear potential based on NFW profile~(orange) and $N$-body simulation~(green).
The black circles with errorbars are the result of $N$-body simulation.
}
\label{fig: dipole Potential large}
\includegraphics[width=0.8\textwidth]{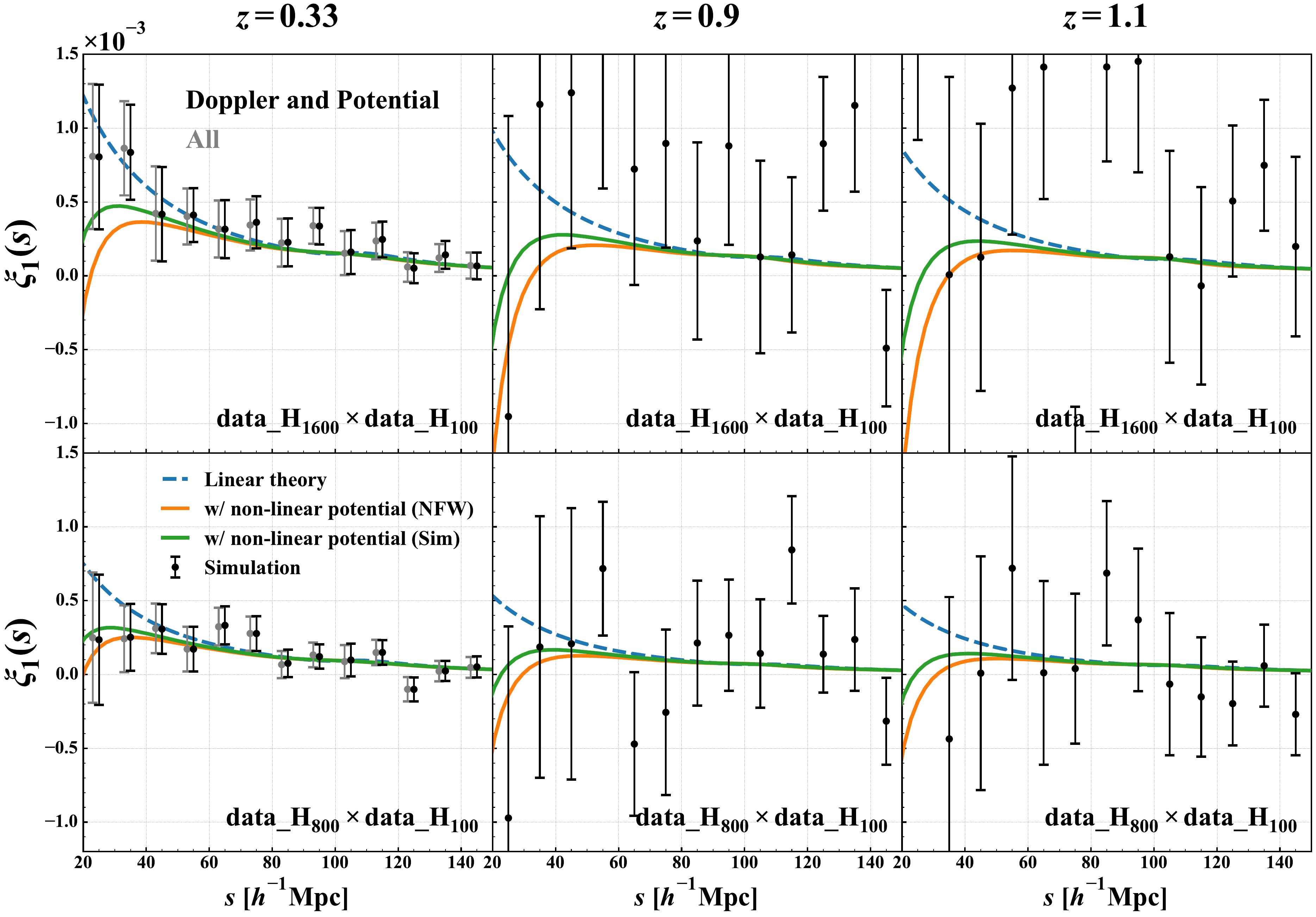}
\caption{Same as Fig.~\ref{fig: dipole Potential large} but for the Doppler and gravitational redshift contributions to the dipole cross-correlation.
In the left panels, we show the dipole moment including all the relativistic effects in $N$-body simulations (grey circle with errorbars).
}
\label{fig: dipole Doppler and Potential large}
\end{figure*}

Let us first look at the dipole cross-correlation at large scales.
Figs.~\ref{fig: dipole Potential large} and \ref{fig: dipole Doppler and Potential large} summarize the results of the comparison at redshifts $z=0.33$ (left), $0.9$ (middle) and $1.1$ (right), focusing on the scales of $20\; h^{-1}{\rm Mpc}\leq s\leq 150\; h^{-1}{\rm Mpc}$.
Here, the cross-correlations between the halo subsample \data{100} and massive counterparts \data{800} and \data{1600} are particularly shown.
Fig.~\ref{fig: dipole Potential large} presents the results including the gravitational redshift effect, while Fig.~\ref{fig: dipole Doppler and Potential large} plots the results including both the gravitational redshift and Doppler effects.

Since the scales we are looking at mostly lies at the linear regime, we do not see any significant difference between linear and quasi-linear predictions especially at $s\gtrsim 60\,h^{-1}{\rm Mpc}$, where the measured dipoles are all in good agreement with the linear theory predictions, although the scatter is somewhat large at $z=0.9$ and $1.1$.
This large scatter at higher redshifts is mainly attributed to the small number of pairs of haloes for cross-correlation (see Table~\ref{tab: population}).

Turning to focus on the scales down to $s\lesssim60\,h^{-1}{\rm Mpc}$, the measured results including only the gravitational redshift effect start to deviate from the linear theory, and they instead tend to follow the quasi-linear predictions.
Though the scatter of the measurement is large at $z=0.9$ and $1.1$, the quasi-linear predictions better explain the simulation results if one adopts the numerically estimated non-perturbative potential (green).
Nevertheless, including both the gravitational redshift and Doppler effects, the difference between linear theory and quasi-linear predictions diminishes, and within the statistical errors, all predictions become mostly consistent with the measured dipoles.
This is because the Doppler effect is the main contributor at the scales shown here, and it gives a positive contribution to the dipole.
For reference, in the left panel of Fig.~\ref{fig: dipole Doppler and Potential large}, we also show the measured dipoles including all other relativistic effects, such as gravitational lensing and integrated Sachs-Wolfe effects.
The results are depicted as filled grey symbols with errorbars, but they remain almost unchanged, compared to those including the Doppler and gravitational redshift effects (black filled symbols).
In this respect, ignoring other relativistic contributions is relevant and validated for a quantitative prediction of dipole cross-correlation.

\subsection{Dipole cross-correlation at small scales}
\label{sec: dipole small scales}

\begin{figure*}
\centering
\includegraphics[width=0.8\textwidth]{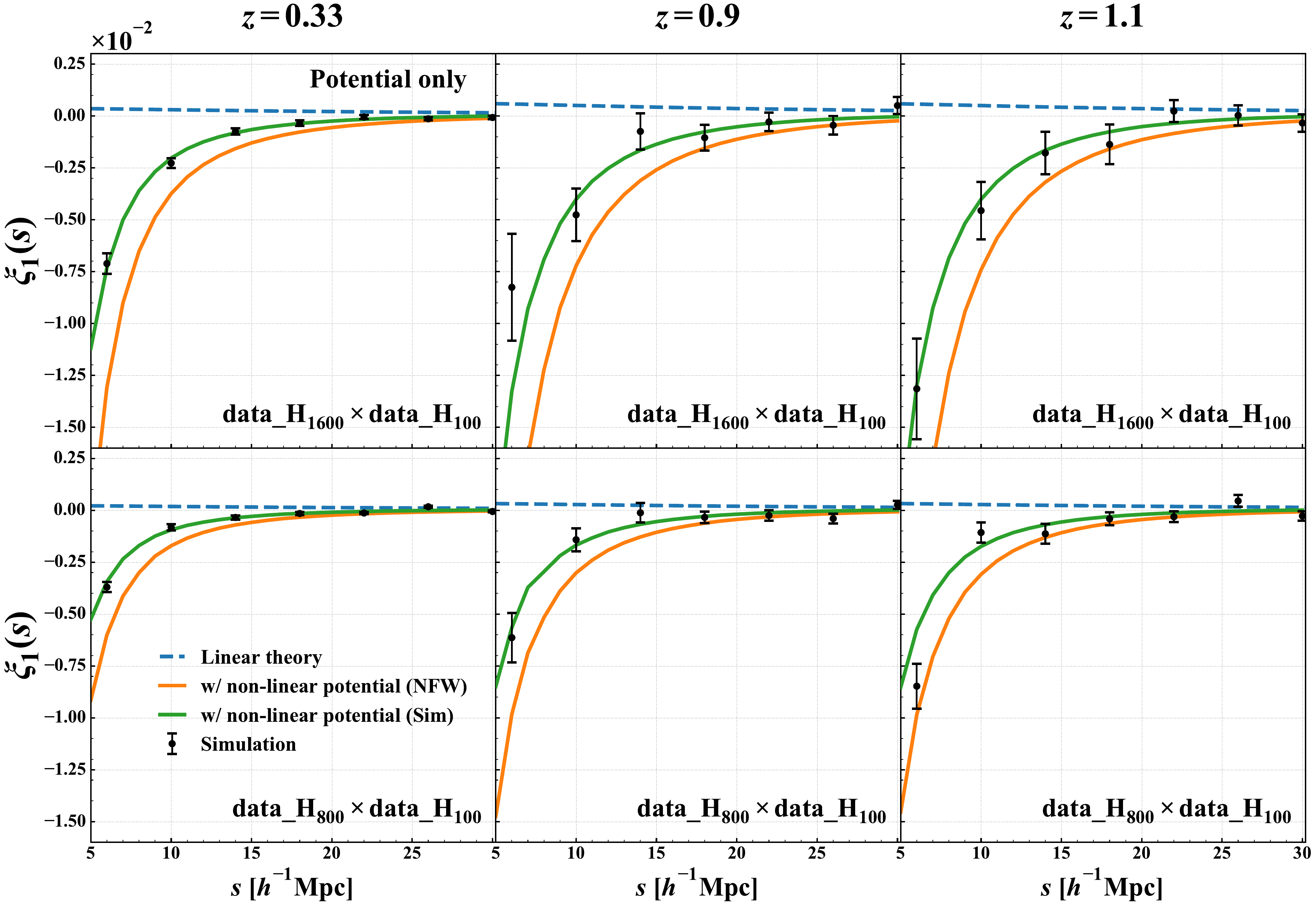}
\caption{Same as Fig.~\ref{fig: dipole Potential large} but for the dipole cross-correlation function at small scales.
}
\label{fig: dipole Potential small}
\includegraphics[width=0.8\textwidth]{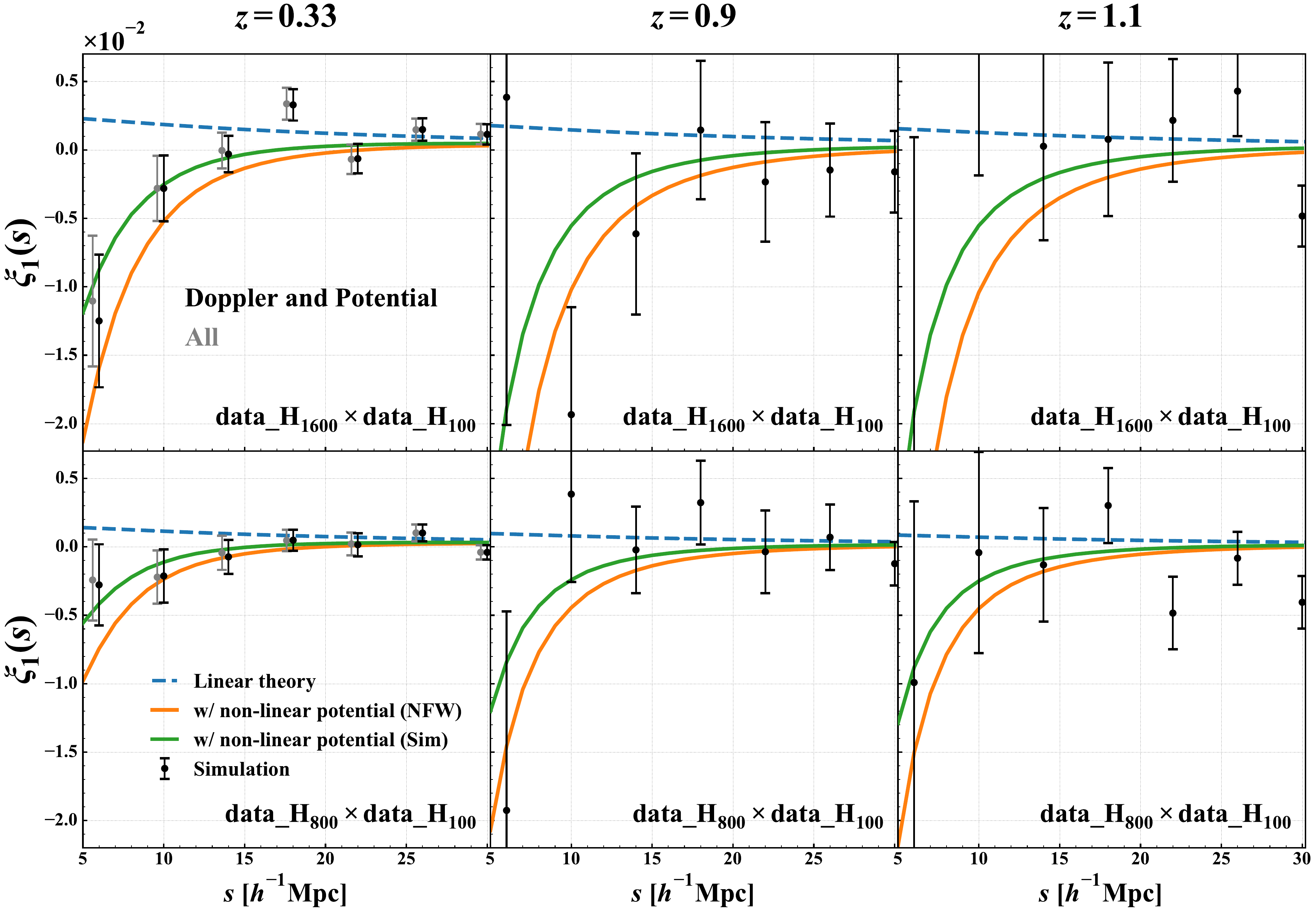}
\caption{Same as Fig.~\ref{fig: dipole Doppler and Potential large} but for the dipole cross-correlation function at small scales.
}
\label{fig: dipole Doppler and Potential small}
\end{figure*}

Next focus on the small-scale behaviors of the dipole cross-correlation functions. Figs.~\ref{fig: dipole Potential small} and \ref{fig: dipole Doppler and Potential small} shows the results at $5\; h^{-1}{\rm Mpc}\leq s\leq 30\; h^{-1}{\rm Mpc}$, with the meanings of the symbols and line types being the same as in Figs.~\ref{fig: dipole Potential large} and \ref{fig: dipole Doppler and Potential large}.

As expected from the analytical predictions shown in Sec.~\ref{sec: bias and redshift dep}, the measured dipoles drop sharply as decreasing the separation $s$, and the deviation from the linear theory prediction, depicted as dashed lines, becomes significant.
The dipole cross-correlation takes the negative value and the quasi-linear predictions explain the overall trends seen in the simulations.
In particular, adopting the non-perturbative halo potential estimated from simulations, the predicted dipoles (green) including only the gravitational redshift effect agree well with simulations (Fig.~\ref{fig: dipole Potential small}), and the agreement is even nice at $s\simeq 5\,h^{-1}{\rm Mpc}$, where the quasi-linear treatment with the Zel'dovich approximation is supposed to become inadequate.
While this can be regarded as a remarkable achievement, a part of the reasons may be ascribed to a large impact of the non-perturbative halo potential at small scales.
Indeed, adopting the NFW profile, the quasi-linear predictions with the non-perturbative potential (orange) fall off more rapidly than the measured dipoles, and the discrepancy becomes manifest at $s\lesssim15\,h^{-1}{\rm Mpc}$.
This indicates that a precision modelling of non-perturbative potential is rather crucial to detect and discriminate the relativistic dipole from others at small scales.

On the other hand, looking at the dipole cross-correlation including both the gravitational redshift and Doppler effects (Fig.~\ref{fig: dipole Doppler and Potential small}), the measured results of the dipoles become rather noisy and their statistical errors get increased at higher redshifts.
Compared to the statistical errors, the differences between the two quasi-linear predictions, depicted as green and orange lines, are rather comparable or small, and both of the predictions reasonably explain the measured trends of the bias and redshift dependences.
Nevertheless, at these scales, the nonlinear cross-talk between the Doppler and gravitational redshift effects becomes important, and a simple superposition of the predictions taking separately each effect into account would fail to reproduce the simulation results.
A proper treatment beyond linear theory is thus crucial, and incorporating the non-perturbative halo potential, our quasi-linear model gives a reasonable description of the dipole cross-correlation at both large and small scales.

\section{Conclusion}
\label{sec: summary}

On top of the well-known contribution from the Doppler effect, the observed galaxy distribution via spectroscopic surveys appears distorted due to the minor but non-negligible contributions coming from the general and special relativistic effects.
Recently, incorporating the relativistic effects arising from the light propagation in an inhomogeneous universe, \citet{2019MNRAS.483.2671B} created a full-sky halo catalogue, and have studied the imprint of such relativistic distortions on the asymmetry of halo clustering.
They found that the dipole cross-correlation (which usually becomes vanishing in the distant-observer limit and can be thus used as a sensible probe of relativistic effects) is dominated by the (standard) Doppler effect at large scales.
However, at small scales, the gravitational redshift effect becomes dominant, and the amplitude of the dipole eventually flip the sign.
The linear theory is unable to describe the behavior of the sign flip, and thus it fails to reproduce the simulation results.

In this paper, we have developed an analytic model of the cross-correlation function beyond linear theory that can explain the numerical results of \citet{2019MNRAS.483.2671B}.
To do so, one crucial aspect is that one needs to take both the relativistic and standard Doppler effects into account, the latter of which is especially known to produce the non-vanishing odd-multipoles through the so-called wide-angle effect.
Here, based on the formalism developed by \citet{2020MNRAS.491.4162T}, we have successfully built a quasi-linear model.
Adopting the Zel'dovich approximation, the model incorporates simultaneously the Doppler and gravitational redshift effects into the prediction of cross-correlation function, in which the wide-angle effect is properly taken into account.
The model also accounts for the non-perturbative contribution to the gravitational redshift effect arising from the halo potential, which is shown to play an important role to describe the small-scale behaviors (i.e., sign flip) of the dipole cross-correlation function.

Provided the (linear) bias parameters and halo potential for the objects to cross-correlate, our quasi-linear model gives a quantitative prediction of the cross-correlation function. In particular, adopting the numerically estimated values of the halo potential, the predictions show a remarkable agreement with the measured dipole cross-correlations, not only at large scales but also at small scales, where the discrepancy between the linear theory and simulations is significant.
We see that the amplitude of the dipoles exhibits a sign flip, and more importantly, the scale at which the sign flip happens is predicted to be shifted to a larger scale as increasing the bias difference and redshift (this matches well the measured trends in simulations).
We have derived a simple approximate formula for the dipole to account for the sign flip (see Eq.~(\ref{eq: explanation negative dipole})).
The prediction based on this formula reproduces quantitatively the results from the full quasi-linear calculations with the non-perturbative corrections, and with this formula, the bias and redshift dependencies on the characteristic scale of the sign flip are qualitatively understood (see Eq.~(\ref{eq:scale_sign_flip})).
Note that our quasi-linear formalism strictly assumes uniform mean number density over the survey region. While this is idealistic, we show that our prediction still works well to explain the simulation results whose halo samples are not strictly uniform across the redshifts. Although, making use of Eq.~(\ref{eq: explanation negative dipole}), it is still possible to incorporate the non-uniform selection function into the predictions, a rigorous quasi-linear treatment is left to future work.

The quasi-linear model presented in this paper has several distinctive features among other recently proposed models for the relativistic dipole.
In particular, the sign flip of the dipole cross-correlation arises from the non-perturbative halo potential. This is close to what has been discussed by \citet{2013MNRAS.434.3008C}, but we incorporate it self-consistently into the quasi-linear prediction, which enables us not only to reproduce the linear theory results at large scales but also to quantitatively explain the simulation results at small scales.
A similar idea has been also advocated by \citet{2017arXiv170907854G}, but they introduced the non-perturbative potential offset by hand in the linear theory prediction, hence unable to reproduce simulation results consistently at both large and small scales.
On the other hand, \citet{2019JCAP...04..050D} have developed the higher-order Eulerian perturbative calculations up to the third order, including the relativistic effects.
Their one-loop (i.e., next-to-leading order) predictions are then shown to agree well with dipole cross-correlations measured by \citet{2019MNRAS.483.2671B}.
A crucial difference of their model from ours is that they introduced the nonlinear bias and effective-field-theory parameters, the latter of which is needed to mitigate the small-scale UV-sensitive behaviors of the Eulerian perturbative treatment.
That is, the sign flip of the dipoles in their model appears manifest through a certain balance between the nonlinear bias and effective-field-theory parameters.
For a more clear understanding of the relativistic dipole at small scales, testing and comparing between the model predictions are therefore important, and we leave it for future work.

Finally, toward a practical application to the observations, one potential concern is that the actual objects to cross-correlate are not the haloes but galaxies.
While the halo is thought to be an ideal formation site of galaxies, depending on the survey specification, the observed galaxies do not fairly trace the halo distribution.
In other words, the location of galaxies does not necessarily correspond to the positions of the deepest potential well in each halo.
Although the analytical model presented here can be applied to the galaxy cross-correlation function, a more careful study is necessary for a solid detection of the gravitational redshift effect, and the impact of the non-perturbative contribution to the gravitational redshift effect especially deserves further investigation.

\section*{Acknowledgments}
This work was initiated during the invitation program of JSPS Grant No. L16519. Numerical simulation was granted access to HPC resources of TGCC through allocations made by GENCI (Grand Equipement National de Calcul Intensif) under the allocations A0030402287 and A0050402287.
Numerical computation was also carried out partly at the Yukawa Institute Computer Facility.
This work was supported by Grant-in-Aid for JSPS Fellows No.~17J10553 (SS) and in part by MEXT/JSPS KAKENHI Grant Numbers JP15H05889 and JP16H03977 (AT).

\section*{Data availability}

The data underlying this article are available in RayGalGroupSims Relativistic Halo Catalogs at \url{https://cosmo.obspm.fr/raygalgroupsims-relativistic-halo-catalogs/}.

\bibliographystyle{mnras}
\bibliography{ref}
\appendix

\section{Derivation of cross-correlation function}
\label{appendix: correlation function}

In this appendix, we present the derivations of the cross-correlation function given in Eqs.~(\ref{eq: RXRY final}) and (\ref{eq: DXDY final}).
The derivation of the expression with the relativistic corrections is a straightforward extension of~\citet{2020MNRAS.491.4162T}.

On the basis of the Zel'dovich approximation, our goal here is to write down the cross-correlation function in terms of the quantities in Lagrangian space. To do so, we first recall the number conservation between redshift space and Lagrangian space. We have 
\begin{equation}
n^{({\rm S})}_{\rm X}(\bm{s}){\rm d}^{3}s =
\overline{n}_{\rm X}
\left( 1 + b^{\rm L}_{\rm X} \delta_{\rm L}(\bm{q})\right){\rm d}^{3}q ~,
\label{eq: number conservation}
\end{equation}
where the quantities $n^{(S)}_{\rm X}(\bm{s})$, $\delta_{\rm L}$, and $\overline{n}_{\rm X}$, respectively, stand for the number density field of the biased objects X in redshift space, the mean number density at a given redshift, and (Lagrangian) linear density field.
The mean number density $\overline{n}_{\rm X}$ is assumed to be constant over the survey region.
Here, the quantity $b^{\rm L}_{\rm X}$ for the population ${\rm X}$ is the Lagrangian linear bias, which is related to the Eulerian linear bias $b_{\rm X}$ through $b_{\rm X}=1+b_{\rm X}^{\rm L}$.

Using Eq.~(\ref{eq: number conservation}) and Eq.~(\ref{eq: mapping_s_q-space epsilon}), the  redshift-space number density, $n^{({\rm S})}_{\rm X}$, is rewritten with~\citep{2020MNRAS.491.4162T}
\begin{align}
n^{({\rm S})}_{\rm X}(\bm{s})
&=
\overline{n}_{\rm X}
\left( 1 + b^{\rm L}_{\rm X} \delta_{\rm L}(\bm{q})\right) \left| \frac{\partial s_{i}}{\partial q_{j}}\right|^{-1}
\notag \\
&=
\overline{n}_{\rm X}
\int{\rm d}^{3}q\;
\Bigl( 1 + b^{\rm L}_{\rm X}\delta_{\rm L}(\bm{q})\Bigr)
\delta^{3}_{\rm D}\left( \bm{s} - \bm{q}
- \bm{\Psi}^{({\rm S})}
- \epsilon_{\rm NL, X} \hat{\bm{q}}
\right) \notag \\
&=
\overline{n}_{\rm X}
\int{\rm d}^{3}q\; 
\int\frac{{\rm d}^{3}k}{(2\pi)^{3}}
\Bigl( 1 + b^{\rm L}_{\rm X}\delta_{\rm L}(\bm{q})\Bigr)
\notag \\
& ~~~ \times 
{\rm e}^{{\rm i}\bm{k}\cdot
\left(\bm{s} - \bm{q}
- \bm{\Psi}^{({\rm S})}
- \epsilon_{\rm NL, X} \hat{\bm{q}}
\right)} ~. \label{eq: RSD n bar}
\end{align}

Then, we define the density fluctuations as
\begin{equation}
\delta^{(\rm S)}_{\rm X}(\bm{s}) \equiv \frac{n^{(S)}_{\rm X}(\bm{s})}{\Braket{n^{(S)}_{\rm X}(\bm{s})}} - 1~,
\end{equation}
where the bracket $\Braket{\cdots}$ implies an ensemble average.
The two-point cross-correlation function between the objects X at $\bm{s}_{1}$ and Y at $\bm{s}_{2}$ in redshift space is given by
\begin{align}
1 + \xi^{(\rm S)}_{\rm XY}(\bm{s}_{1},\bm{s}_{2})
&= \Braket{\left( 1+\delta^{(\rm S)}_{\rm X}(\bm{s}_{1}) \right) \left( 1 + \delta^{(\rm S)}_{\rm Y}(\bm{s}_{2})\right) }
\notag \\
&\equiv \frac{\DD(\bm{s}_{1},\bm{s}_{2})}{\RX(\bm{s}_{1}) \RY(\bm{s}_{2})} ~,
\label{eq: xi def app}
\end{align}
where the functions $\RX$ (or $\RY$) and $\DD$ are respectively given by Eqs.~(\ref{eq: RX braket}) and (\ref{eq: DD}). 
In deriving the explicit expression for $\RX$, $\RY$ below, and $\DD$ below, it is convenient to define the following statistical quantities in Lagrangian space:
\begin{align}
A_{ij}(\hat{\bm{q}}) &= \Braket{\Psi^{(\rm S)}_{i}(\bm{q})\Psi^{(\rm S)}_{j}(\bm{q})}_{\rm c} ~, \label{eq: Aij def}\\
y_{i}(\hat{\bm{q}}) &= \Braket{\delta_{\rm L}(\bm{q})\Psi^{(\rm S)}_{i}(\bm{q})}_{\rm c} ~, \label{eq: yi def}\\
B_{ij}(\bm{q}_{1},\bm{q}_{2}) &= \Braket{\Psi^{(\rm S)}_{i}(\bm{q}_{1})\Psi^{(\rm S)}_{j}(\bm{q}_{2})}_{\rm c} ~, \\
U_{i}(\bm{q}_{1},\bm{q}_{2}) &= - \Braket{\delta_{\rm L}(\bm{q}_{1})\Psi^{(\rm S)}_{i}(\bm{q}_{2})}_{\rm c} ~, \label{eq: Ui def}
\end{align}
where the field $\bm{\Psi}^{({\rm S})}$ is defined at  Eq.~(\ref{eq: Psi RSD}).
The bracket $\Braket{\cdots}_{\rm c}$ represents the cumulant for which the disconnected part is subtracted.
Here, the fields, $\bm{\Psi}(\bm{q})$ and $\epsilon_{\rm L}(\bm{q})$, are assumed to follow the Gaussian statistics.

Taking advantage of their symmetric properties, 
the expressions given at Eqs.~(\ref{eq: Aij def})--(\ref{eq: Ui def}) are rewritten as follows: 
\begin{align}
A_{ij}(\hat{\bm{q}}) &= R_{ik}(\hat{\bm{q}})R_{jk}(\hat{\bm{q}})\sigma^{2}_{\rm d} + \hat{q}_{i}\hat{q}_{j}\sigma^{2}_{\epsilon} ~, \label{eq: Aij def 2}\\
y_{i}(\hat{\bm{q}}) &= \hat{q}_{i}\sigma^{2}_{\delta\epsilon} ~, \label{eq: def y i}\\
B_{ij}(\bm{q}_{1},\bm{q}_{2}) &= R_{ik}(\hat{\bm{q}}_{1}) R_{jl}(\hat{\bm{q}}_{2}) \Bigl( C(q)\delta_{ij} + D(q)\hat{q}_{i}\hat{q}_{j}\Bigr)
\notag \\
&
+ \hat{q}_{1i}\hat{q}_{2j}\xi_{\epsilon}(q)
+ \Bigl( R_{ik}(\hat{\bm{q}}_{1})\hat{q}_{2j} - R_{jk}(\hat{\bm{q}}_{2})\hat{q}_{1i}\Bigr) \hat{q}_{k} L_{\epsilon}(q) ~, \\
U_{i}(\bm{q}_{1},\bm{q}_{2}) &= R_{ij}(\hat{\bm{q}}_{2})\hat{q}_{j}L_{\delta}(q) - \hat{q}_{2i}\xi_{\delta\epsilon}(q) ~, \label{eq: Ui def 2}
\end{align}
with $\bm{q} = \bm{q}_{2} - \bm{q}_{1}$ and $q \equiv |\bm{q}|$.
Here we define
\begin{align}
\sigma^{2}_{\rm d}\delta_{ij} &= \Braket{\Psi_{i}(\bm{q})\Psi_{j}(\bm{q})}_{\rm c}~, \\ 
\sigma^{2}_{\epsilon} &= \Braket{\epsilon_{\rm L}(\bm{q})\epsilon_{\rm L}(\bm{q})}_{\rm c} ~, \\
\sigma^{2}_{\delta\epsilon} &= \Braket{\delta_{\rm L}(\bm{q})\epsilon_{\rm L}(\bm{q})}_{\rm c} ~, \\
C(q)\delta_{ij} + D(q)\hat{q}_{i}\hat{q}_{j} &= \Braket{\Psi_{i}(\bm{q}_{1})\Psi_{j}(\bm{q}_{2})}_{\rm c}~, \\
\xi_{\rm L}(q) &= \Braket{\delta_{\rm L}(\bm{q}_{1})\delta_{\rm L}(\bm{q}_{2})}_{\rm c}~, \label{eq: def xi_L}\\
\xi_{\epsilon}(q) &= \Braket{\epsilon_{\rm L}(\bm{q}_{1})\epsilon_{\rm L}(\bm{q}_{2})}_{\rm c} ~, \\
\xi_{\delta\epsilon}(q) &= \Braket{\delta_{\rm L}(\bm{q}_{1})\epsilon_{\rm L}(\bm{q}_{2})}_{\rm c} ~, \\
L_{\epsilon}(q)\hat{q}_{i} &= \Braket{\Psi_{i}(\bm{q}_{1})\epsilon_{\rm L}(\bm{q}_{2})}_{\rm c}~, \\
L_{\delta}(q)\hat{q}_{i} &= \Braket{\Psi_{i}(\bm{q}_{1})\delta_{\rm L}(\bm{q}_{2})}_{\rm c}~.
\end{align}
Using the fact that the displacement field $\bm{\Psi}$ is related to the linear density field through 
$\bm{\nabla}\cdot\bm{\Psi} (\bm{q},t) = -\delta_{\rm L}(\bm{q},t)$ (see Eq.~(\ref{eq:psi_to_delta_ZA}), we obtain
\begin{align}
\sigma^{2}_{\rm d} &= \frac{1}{3}\int{\frac{{\rm d}p}{2\pi^{2}}}\; P_{\rm L}(p) ~, \label{eq: def sigma2d}\\ 
\sigma^{2}_{\epsilon} &= \int{\frac{{\rm d}p}{2\pi^{2}}}\; p^{2}P_{\epsilon}(p) ~, \label{eq: def sigma2e}\\
\sigma^{2}_{\delta\epsilon} &= \int{\frac{{\rm d}p}{2\pi^{2}}}\; p^{2}P_{\delta \epsilon}(p) ~, \label{eq: def sigma2de}\\
C(q)&= \int{\frac{{\rm d}p}{2\pi^{2}}}\; \frac{j_{1}(pq)}{pq}P_{\rm L}(p) ~, \\
D(q)&= - \int{\frac{{\rm d}p}{2\pi^{2}}}\; j_{2}(pq) P_{\rm L}(p) ~, \\
\xi_{\rm L}(q) &= \int{\frac{{\rm d}p}{2\pi^{2}}}\; p^{2} j_{0}(pq) P_{\rm L}(p) ~, \\
\xi_{\epsilon}(q) &= \int{\frac{{\rm d}p}{2\pi^{2}}}\; p^{2} j_{0}(pq) P_{\epsilon}(p) ~, \\
\xi_{\delta\epsilon}(q) &= \int{\frac{{\rm d}p}{2\pi^{2}}}\; p^{2} j_{0}(pq) P_{\delta\epsilon}(p) ~, \\
L_{\epsilon}(q) &= \int{\frac{{\rm d}p}{2\pi^{2}}}\; p j_{1}(pq) P_{\delta\epsilon}(p) ~, \\
L_{\delta}(q) &= \int{\frac{{\rm d}p}{2\pi^{2}}}\; p j_{1}(pq) P_{\rm L}(p) ~,
\end{align}
where the spectra $P_{\rm L}$, $P_{\delta\epsilon}$, and $P_{\epsilon\epsilon}$ are defined as follows:
\begin{align}
\Braket{\delta_{\rm L}(\bm{k})\delta_{\rm L}(\bm{k}')}_{\rm c} &= (2\pi)^{3}\delta^{3}_{\rm D}(\bm{k}+\bm{k}') P_{\rm L}(k) ~, \\
\Braket{\delta_{\rm L}(\bm{k})\epsilon_{\rm L}(\bm{k}')}_{\rm c} &= (2\pi)^{3}\delta^{3}_{\rm D}(\bm{k}+\bm{k}') P_{\delta\epsilon}(k) ~, \\
\Braket{\epsilon_{\rm L}(\bm{k})\epsilon_{\rm L}(\bm{k}')}_{\rm c} &= (2\pi)^{3}\delta^{3}_{\rm D}(\bm{k}+\bm{k}') P_{\epsilon}(k) ~.
\end{align}

With the quantities given above, we will derive an analytical expression for $\RX$ (or $\RY$) in Sec.~\ref{appendix: RX RY}, and for $\DD$ in Sec.~\ref{appendix: DD}.

\subsection{$\RX$ and $\RY$ part}
\label{appendix: RX RY}

In order to a simplified analytical expressions of $\RX$ (or $\RY$), we use the properties between moment and cumulant generating function~(see e.g.,~\citealt{2004PhRvD..70h3007S,2008PhRvD..77f3530M,2010PhRvD..82f3522T}). The integrand of  Eq.~(\ref{eq: RXRY final}) is then rewritten in the form as
\begin{align}
&
\Braket{
\left( 1 + b^{\rm L}_{\rm X}\delta_{\rm L}(\bm{q})\right)
{\rm e}^{- {\rm i}\bm{k}\cdot \bm{\Psi}^{(\rm S)}(\bm{q})}
}
\notag \\
&= \exp\left[ -\frac{1}{2}k_{i}k_{j}A_{ij}\right]
\left( 1 - {\rm i}b^{\rm L}_{\rm X}k_{i}y_{i}\right) ~,
\label{eq: braket RX RY}
\end{align}
where the functions $A_{ij}$and $y_{i}$ are defined in Eqs.~(\ref{eq: Aij def 2}) and (\ref{eq: def y i}), respectively.
Substituting Eq.~(\ref{eq: braket RX RY}) into Eq.~(\ref{eq: RX braket}), we obtain 
\begin{align}
\RX(\bm{s}_{1})	
&=
\int{\rm d}^{3}q\; 
\int\frac{{\rm d}^{3}k}{(2\pi)^{3}}\;
{\rm e}^{{\rm i}\bm{k}\cdot \left( \bm{s}_{1} - \bm{q} - \epsilon_{\rm NL, X} \hat{\bm{q}}\right)}
\notag \\
&\times 
\exp\left[ -\frac{1}{2}k_{i}k_{j}A_{ij}\right]
\left( 1 - {\rm i}b^{\rm L}_{\rm X}k_{i}y_{i}\right)
\label{eq: RX full app} \\
&= 
\int\frac{{\rm d}^{3}q}{(2\pi)^{3/2}|\det{A}|^{1/2}}
\notag \\
&\times 
{\rm e}^{-\frac{1}{2}\left( s_{1i}-q_{i}- \epsilon_{\rm NL, X} \hat{q}_{i} \right)\left( s_{1j}-q_{j}- \epsilon_{\rm NL, X} \hat{q}_{j} \right)
A^{-1}_{ij}}
\notag \\
&\times 
\left( 1+ b^{\rm L}_{\rm X}\left( s_{1j}-q_{j}- \epsilon_{\rm NL, X} \hat{q}_{j}\right)A^{-1}_{ij}y_{i}\right) ~.
\end{align}
In the second line, we used the formula for multi-dimensional Gaussian integrals to obtain the second equality.
The final expression of $\RX$ (or $\RY$) involves the three-dimensional integral, which has to be computed numerically.  

Nevertheless, we can exploit the approximation, with which the $\RX$ (or $\RY$) is expressed in a simple analytical form. Following \citet{2020MNRAS.491.4162T}
(see their Appendix A1), we Taylor-expand the exponential factor in Eq.~(\ref{eq: RX full app}),  and repeat the integration by part. Truncating at finite order, the final form becomes 
\begin{align}
\RX&(s) =
1 - \frac{2}{s}b^{\rm L}_{\rm X} \sigma^{2}_{\delta\epsilon}
\notag \\
&
+ \left( \frac{\sigma^{2}_{\rm d}}{s^{2}} + \frac{\sigma^{4}_{\rm d}}{s^{4}} + 3 \frac{\sigma^{6}_{\rm d}}{s^{6}} + \cdots \right)\left( 2f + f^{2} + \frac{\sigma^{2}_{\epsilon}}{\sigma^{2}_{\rm d}}\right) \notag \\
& + \frac{2}{s}\epsilon_{\rm NL, X}
\notag \\
& \times 
\Biggl[ - 1 + \frac{1}{s}b^{\rm L}_{\rm X} \sigma^{2}_{\delta\epsilon} + \left( \frac{\sigma^{4}_{\rm d}}{s^{4}}+ 6\frac{\sigma^{6}_{\rm d}}{s^{6}} + \cdots \right) \left( 2f + f^{2} + \frac{\sigma^{2}_{\epsilon}}{\sigma^{2}_{\rm d}}\right)\Biggr] ~,
\end{align}
which is accurate as long as we consider the cases with $\sigma_{\rm d}^2/s^2\ll1$, $\sigma_{\delta\epsilon}^2/s\ll1$, and $\sigma_\epsilon^2/\sigma_{\rm d}^2\sim\mathcal{O}(1)$.

\subsection{$\DD$ part}
\label{appendix: DD}

Next consider the function $\DD$, defined at Eq.~(\ref{eq: DD}). To derive a simplified analytical expression, we first define the following quantities:
\begin{align}
Y &= 
- {\rm i} \left( \bm{k}_{1}\cdot \bm{\Psi}^{(\rm S)}(\bm{q}_{1})
+ \bm{k}_{2}\cdot \bm{\Psi}^{(\rm S)}(\bm{q}_{2}) \right) ~, \\
X_{1} &= b^{\rm L}_{\rm X}\delta_{\rm L}(\bm{q}_{1}) ~, \\
X_{2} &= b^{\rm L}_{\rm Y}\delta_{\rm L}(\bm{q}_{2}) ~.
\end{align}
The quantities $X_{1}$, $X_{2}$, and $Y$ follow the Gaussian statistics.
Then, Eq.~(\ref{eq: DD}) is expressed in terms of $Y$, $X_{1}$, and $X_{2}$ as follows:
\begin{align}
\DD(\bm{s}_{1},\bm{s}_{2})
&= 
\int{\rm d}^{3}q_{1}
\int{\rm d}^{3}q_{2}
\int\frac{{\rm d}^{3}k_{1}}{(2\pi)^{3}}
\int\frac{{\rm d}^{3}k_{2}}{(2\pi)^{3}}
\notag \\
&\times 
{\rm e}^{{\rm i}\bm{k}_{1}\cdot \left( \bm{s}_{1} - \bm{q}_{1} - \epsilon_{\rm NL, X} \hat{\bm{q}}_{1}\right)
+{\rm i} \bm{k}_{2}\cdot \left( \bm{s}_{2} - \bm{q}_{2} - \epsilon_{\rm NL, Y} \hat{\bm{q}}_{2}\right)}
\notag \\
&\times 
\Biggl< {\rm e}^{Y}\left( 1 + X_{1}\right)\left( 1 + X_{2}\right) \Biggr> ~.
\label{eq: DD X1 X2 Y}
\end{align}
In similar manner to $\RX$ in Sec.~\ref{appendix: RX RY}, we rewrite the ensemble average of Eq.~(\ref{eq: DD X1 X2 Y}) in terms of the cumulants: 
\begin{align}
\Biggl<
{\rm e}^{Y} \left( 1 + X_{1}\right) & \left( 1 + X_{2}\right)
\Biggr>
= 
{\rm e}^{\frac{1}{2}\Braket{Y^{2}}_{\rm c}}
\Bigl( 1 + \Braket{X_{1}X_{2}}_{\rm c}
\notag \\
& + \Braket{X_{1}Y}_{\rm c} + \Braket{X_{2}Y}_{\rm c} + \Braket{X_{1}Y}_{\rm c}\Braket{X_{2}Y}_{\rm c}\Bigr) ~.
\end{align}
Using Eqs.~(\ref{eq: Aij def})--(\ref{eq: Ui def}), the statistical quantities at right-hand side are explicitly given as follows:
\begin{align}
\Braket{Y^{2}}_{\rm c} &= -k_{1i}k_{1j}A_{ij}(\hat{\bm{q}}_{1}) - k_{2i}k_{2j}A_{ij}(\hat{\bm{q}}_{2})
\notag \\
& - 2k_{1i}k_{2j}B_{ij}(\bm{q}_{1},\bm{q}_{2}) ~, 
\\
\braket{X_{1}Y}_{\rm c} &= {\rm i} b^{\rm L}_{\rm X} \Bigl( k_{2i}U_{i}(\bm{q}_{1},\bm{q}_{2}) - k_{1i}y_{i}(\hat{\bm{q}}_{1}) \Bigr) ~, 
\\
\braket{X_{2}Y}_{\rm c} &= i b^{\rm L}_{\rm Y} \Bigl( k_{1i}U_{i}(\bm{q}_{2},\bm{q}_{1}) -k_{2i}y_{i}(\hat{\bm{q}}_{2}) \Bigr) ~, 
\\
\braket{X_{1}X_{2}}_{\rm c} &= b^{\rm L}_{\rm X}b^{\rm L}_{\rm Y}\xi_{\rm L}(q) ~.
\end{align}
Substituting these expressions into Eq.~(\ref{eq: DD X1 X2 Y}), the correlator $\DD$ is expressed as
\begin{align}
&\DD(\bm{s}_{1},\bm{s}_{2})
= 
\int{\rm d}^{3}q_{1}
\int{\rm d}^{3}q_{2}
\int\frac{{\rm d}^{3}k_{1}}{(2\pi)^{3}}
\int\frac{{\rm d}^{3}k_{2}}{(2\pi)^{3}}
\notag \\
&\times 
{\rm e}^{{\rm i}\bm{k}_{1}\cdot \left( \bm{s}_{1} - \bm{q}_{1} - \epsilon_{\rm NL, X} \hat{\bm{q}}_{1}\right)
+{\rm i} \bm{k}_{2}\cdot \left( \bm{s}_{2} - \bm{q}_{2} - \epsilon_{\rm NL, Y} \hat{\bm{q}}_{2}\right)}
\notag \\
&\times 
\exp\Biggl[
-\frac{1}{2}k_{1i}k_{1j}A_{ij}(\hat{\bm{q}}_{1})
-\frac{1}{2}k_{2i}k_{2j}A_{ij}(\hat{\bm{q}}_{2})
- k_{1i}k_{2j}B_{ij}(\bm{q}_{1},\bm{q}_{2})
\Biggr]
\notag \\
& \times 
\Biggl[
1 + b^{\rm L}_{\rm X}b^{\rm L}_{\rm Y}\xi_{\rm L}(q)
+ {\rm i} k_{2i} \left( b^{\rm L}_{\rm X}U_{i}(\bm{q}_{1},\bm{q}_{2})- b^{\rm L}_{\rm Y}y_{i}(\hat{\bm{q}}_{2}) \right)
\notag \\
&
+ {\rm i} k_{1i} \left( b^{\rm L}_{\rm Y}U_{i}(\bm{q}_{2},\bm{q}_{1}) - b^{\rm L}_{\rm X} y_{i}(\hat{\bm{q}}_{1}) \right)
\notag \\
&
- k_{1i}k_{2j} b^{\rm L}_{\rm X}b^{\rm L}_{\rm Y}
\Bigl( U_{i}(\bm{q}_{2},\bm{q}_{1})U_{j}(\bm{q}_{1},\bm{q}_{2}) + y_{i}(\hat{\bm{q}}_{1})y_{j}(\hat{\bm{q}}_{2})\Bigr)
\notag \\
&
+ k_{1i}k_{1j}b^{\rm L}_{\rm X}b^{\rm L}_{\rm Y}U_{i}(\bm{q}_{2},\bm{q}_{1})y_{j}(\hat{\bm{q}}_{1})
+ k_{2i}k_{2j}b^{\rm L}_{\rm X}b^{\rm L}_{\rm Y}U_{i}(\bm{q}_{1},\bm{q}_{2})y_{j}(\hat{\bm{q}}_{2})
\Biggr] ~.
\label{eq: DXDY full}
\end{align}
Note that setting $\epsilon_{\rm L} = \epsilon_{\rm NL, X} = \epsilon_{\rm NL, Y} = 0$, Eq.~(\ref{eq: DXDY full}) coincides with the expression obtained by~\citet{2020MNRAS.491.4162T}.

For further reduction, we introduce the six-dimensional vectors composed of two three-dimensional vectors as
$\bm{K} \equiv (\bm{k}_{1},\bm{k}_{2})$, $\bm{Q} \equiv (\bm{q}_{1}, \bm{q}_{2})$, $\bm{S} \equiv (\bm{s}_{1}, \bm{s}_{2})$, and $\bm{E} \equiv \left( \epsilon_{\rm NL, X} \hat{\bm{q}}_{1}, \epsilon_{\rm NL, Y} \hat{\bm{q}}_{2} \right)$. Then Eq.~(\ref{eq: DXDY full}) is simplified as follows:
\begin{align}
\DD(\bm{s}_{1},\bm{s}_{2})
& = 
\int{\rm d}^{6}Q
\int\frac{{\rm d}^{6}K}{(2\pi)^{6}}\; 
{\rm e}^{{\rm i}\bm{K}\cdot \left( \bm{S}-\bm{Q}-\bm{E}\right) }
{\rm e}^{-\frac{1}{2}K_{a}K_{b}\mathcal{A}_{ab}(\bm{Q})}
\notag \\
& \times 
\Biggr(
1 + b^{\rm L}_{\rm X}b^{\rm L}_{\rm Y}\xi_{\rm L}(q)
+ {\rm i}K_{a}\mathcal{U}_{a}(\bm{Q})
- K_{a}K_{b}\mathcal{W}_{ab}(\bm{Q})
\Biggl) ~,
\label{eq: DD six*2 integral}
\end{align}
where the subscripts $a$ and $b$ run over 1--6.
The six-dimensional vector $\mathcal{U}$ and matrices $\mathcal{A}$ and $\mathcal{W}$ are defined by:
\begin{align}
\mathcal{A} &=
\left( 
\begin{array}{c c}
\bm{A}(\bm{q}_{1}) & \bm{B}(\bm{q}_{1},\bm{q}_{2}) \\
{}^{\rm T}\bm{B}(\bm{q}_{1},\bm{q}_{2}) & \bm{A}(\bm{q}_{2})
\end{array}
\right) 
~, \label{eq: def mathcal A}\\
\mathcal{W} &=
\frac{1}{2}b^{\rm L}_{\rm X}b^{\rm L}_{\rm Y}
 \left(
\begin{array}{c c}
\bm{W}_{1}(\bm{q}_{1},\bm{q}_{2})
&
\bm{W}_{2}(\bm{q}_{1},\bm{q}_{2})
\\
\bm{W}_{2}(\bm{q}_{2},\bm{q}_{1})
& 
\bm{W}_{1}(\bm{q}_{2},\bm{q}_{1})
\end{array}
\right)
~, \label{eq: def mathcal W} \\
\mathcal{U} &=
\left(
\begin{array}{c}
b^{\rm L}_{\rm Y}\bm{U}(\bm{q}_{2},\bm{q}_{1}) - b^{\rm L}_{\rm X}\bm{y}(\hat{\bm{q}}_{1}) \\
b^{\rm L}_{\rm X}\bm{U}(\bm{q}_{1},\bm{q}_{2}) - b^{\rm L}_{\rm Y}\bm{y}(\hat{\bm{q}}_{2})
\end{array}
\right) ~,\label{eq: def mathcal U}
\end{align}
with the three-dimensional matrices given by
\begin{align}
W_{1,ij}(\bm{q}_{1},\bm{q}_{2}) &= -\left( U_{i}(\bm{q}_{2},\bm{q}_{1})y_{j}(\hat{\bm{q}}_{1}) + y_{i}(\hat{\bm{q}}_{1})U_{j}(\bm{q}_{2},\bm{q}_{1})\right) ~, \\
W_{2,ij}(\bm{q}_{1},\bm{q}_{2}) &= U_{i}(\bm{q}_{2},\bm{q}_{1})U_{j}(\bm{q}_{1},\bm{q}_{2}) + y_{i}(\hat{\bm{q}}_{1})y_{j}(\hat{\bm{q}}_{2}) ~.
\end{align}
These matrices and vectors include the relativistic effects on top of the standard Doppler effect as an extension of the expression obtained by~\citet{2020MNRAS.491.4162T}.
Using the formulae for the multi-dimensional Gaussian integrals, the integral over the six-dimensional wavevector $\bm{K}$ in Eq.~(\ref{eq: DD six*2 integral}) is analytically performed, and we obtain
\begin{align}
& \DD(\bm{s}_{1},\bm{s}_{2})
\notag \\
&= 
\int\frac{{\rm d}^{6}Q}{(2\pi)^{3}|\det{\mathcal{A}}|^{1/2}}
{\rm e}^{-\frac{1}{2}(S-Q-E)_{a}(S-Q-E)_{b}\mathcal{A}^{-1}_{ab}} \notag \\
&\times
\Biggl[
1 + b^{\rm L}_{\rm X}b^{\rm L}_{\rm Y}\xi_{\rm L}(q)
-\mathcal{A}^{-1}_{ab}\mathcal{U}_{a}(S-Q-E)_{b}
\notag \\
&
- \Bigl( \mathcal{A}^{-1}_{ab}-\mathcal{A}^{-1}_{ac}\mathcal{A}^{-1}_{bd}(S-Q-E)_{c}(S-Q-E)_{d}
\Bigr)\mathcal{W}_{ab}
\Biggr] ~.
\end{align}
This is Eq.~(\ref{eq: DXDY final}).
The numerator $\DD$ is expressed by the six-dimensional integral of $\bm{Q}$, which we have to evaluate  numerically.

\section{Linear theory limit of the quasi-linear model} 
\label{appendix: linearization}

In this appendix, we show that our quasi-linear model of cross-correlation function consistently reproduces the expressions of the linear theory with the gravitational redshift effect previously derived in the literature \citep[e.g.,][]{2009JCAP...11..026M,2014PhRvD..89h3535B}.
The explicit form of the gravitational redshift effect is given by 
\begin{equation}
\epsilon_{\rm L}(\bm{x}) = -\frac{\phi_{\rm L}(\bm{x})}{aH} ~. \label{eq: potential app}
\end{equation}
The gravitational potential $\phi$ in Fourier space is related to the linear density field through the Poisson equation as [see Eq.~(\ref{eq: Poisson eq})]
\begin{equation}
\phi_{\rm L}(\bm{k}) = -\frac{3\Omega_{\rm m0}H^{2}_{0}}{2a}\frac{\delta_{\rm L}(\bm{k})}{k^{2}} ~. \label{eq: Fourier Poisson}
\end{equation}

Below, taking into account both the wide-angle and gravitational redshift effects, we show in Sec.~\ref{appendix: density field level} that 
the linear density field is reconstructed from our quasi-linear model with Zel'dovich approximation.  Further, in Sec.\ref{appendix: correlation function level}, starting with the cross-correlation function at Eq.~(\ref{eq: def xi XY}), the linear-order correlation function is shown to be recovered.  

\subsection{Linearized density field}
\label{appendix: density field level}

In this subsection, we focus on the linearized density contrast.
In Eq.~(\ref{eq: RSD n bar}), expanding the terms up to $O(\delta_{\rm L})$, $O(\bm{\Psi}^{(\rm S)})$, and $O(\epsilon_{\rm NL,X})$, the number density field in redshift space and its ensemble average are given by
\begin{align}
n^{(\rm S)}_{\rm X}(\bm{s}) &\simeq
\overline{n}_{\rm X} \int{\rm d}^{3}q\; 
\int\frac{{\rm d}^{3}k}{(2\pi)^{3}}\notag \\
&\times 
\Bigl(
1 + b^{\rm L}_{\rm X}\delta_{\rm L}(\bm{q})
- {\rm i}\bm{k}\cdot\bm{\Psi}^{({\rm S})}
- \epsilon_{\rm NL, X} {\rm i}\bm{k}\cdot\hat{\bm{q}}
\Bigr)
{\rm e}^{{\rm i}\bm{k}\cdot \left(\bm{s} - \bm{q} \right)} ~, \\
\Braket{n^{(\rm S)}_{\rm X}(\bm{s})} &\simeq
\overline{n}_{\rm X} \int{\rm d}^{3}q\; 
\int\frac{{\rm d}^{3}k}{(2\pi)^{3}}
\Bigl(
1 - \epsilon_{\rm NL, X} {\rm i}\bm{k}\cdot\hat{\bm{q}}
\Bigr)
{\rm e}^{{\rm i}\bm{k}\cdot \left(\bm{s} - \bm{q} \right)} ~.
\label{eq: braket n = bar n}
\end{align}
Using these expressions, we obtain the linear density contrast in redshift space as
\begin{align}
\delta^{(\rm S)}_{\rm X}(\bm{s})
& = \frac{n^{({\rm S})}_{\rm X}(\bm{s})}{\Braket{n^{({\rm S})}_{\rm X}(\bm{s})}} - 1
\notag \\
& \simeq
\int{\rm d}^{3}q
\int\frac{{\rm d}^{3}k}{(2\pi)^{3}}
{\rm e}^{{\rm i}\bm{k}\cdot \left(\bm{s} - \bm{q} \right)}
\Bigl( b^{\rm L}_{\rm X}\delta_{\rm L}(\bm{q})
- {\rm i} \bm{k}\cdot\bm{\Psi}^{({\rm S})}
\Bigr)
~,
\label{eq: linearise}
\end{align}
where in the second line, we neglect higher-order terms of the non-linear potential $\epsilon_{\rm NL,X}$.
As seen from Eq.~(\ref{eq: linearise}), at linear order, the non-linear potential term $\epsilon_{\rm NL, X}$ vanishes.

Substituting Eqs.~(\ref{eq: Psi RSD}), (\ref{eq: potential app}), and (\ref{eq: Fourier Poisson}) into Eq.~(\ref{eq: linearise}), we obtain
\begin{align}
\delta^{({\rm S})}_{\rm X}(\bm{s})
&= \int\frac{{\rm d}^{3}k}{(2\pi)^{3}}\; e^{{\rm i}\bm{k}\cdot\bm{s}}
\Biggl[
b_{\rm X} + f\mu^{2} - {\rm i} f \frac{2}{ks} \mu
\notag \\
&
- {\rm i} \frac{3}{2} \frac{\Omega_{\rm m0}H^{2}_{0}}{H} \frac{\mu}{k}
- \frac{3}{2} \frac{\Omega_{\rm m0}H^{2}_{0}}{H}\frac{1}{k^{2}}\frac{2}{s}
\Biggr] \delta_{\rm L}(\bm{k}) ~, \label{eq: linear density final}
\end{align}
where the directional cosine $\mu$ is defined by $\mu = \hat{\bm{k}}\cdot\hat{\bm{s}}$, and we introduce the Eulerian linear bias given by $b_{\rm X} = 1 + b^{\rm L}_{\rm X}$.
Except for the selection function terms that is proportional to $2/s$, the integrand of Eq.~(\ref{eq: linear density final}) consistently reproduces the linear theory result in \citet{2009JCAP...11..026M,2014PhRvD..89h3535B}.

Note that, in Eq.~(\ref{eq: braket n = bar n}), neglecting the non-linear potential term, $\overline{n}_{\rm X}$ coincides with the mean density in redshift space $\langle n^{(\rm S)}_{\rm X}(\bm{s}) \rangle$ at linear level.
However, at non-linear level, we see the discrepancy between $\overline{n}_{\rm X}$ and $\langle n^{(\rm S)}_{\rm X}(\bm{s}) \rangle$ due to the wide-angle and relativistic effects~\citep{2020MNRAS.491.4162T}.

\subsection{Linearized cross-correlation function}
\label{appendix: correlation function level}

Second, we show that the cross-correlation function based on our formalism consistently recovers the linear cross-correlation function presented in~\citet{2009JCAP...11..026M,2014PhRvD..89h3535B}.
In this appendix, the procedure of the calculation technique is based on Appendix~C of~\citet{2020MNRAS.491.4162T}.

In Eq.~(\ref{eq: DXDY full}), we keep and expand the terms up to $O\left( P_{\rm L}(k) \right)$.
Then the linearized expression of $\DD$ in Eq.~(\ref{eq: DXDY full}) is given as
\begin{align}
&\DD(\bm{s}_{1},\bm{s}_{2})
= 
\int{\rm d}^{3}q_{1}
\int{\rm d}^{3}q_{2}
\int\frac{{\rm d}^{3}k_{1}}{(2\pi)^{3}}
\int\frac{{\rm d}^{3}k_{2}}{(2\pi)^{3}}
\notag \\
& \times 
{\rm e}^{{\rm i}\bm{k}_{1}\cdot \left( \bm{s}_{1} - \bm{q}_{1} - \epsilon_{\rm NL, X} \hat{\bm{q}}_{1}\right)
+{\rm i} \bm{k}_{2}\cdot \left( \bm{s}_{2} - \bm{q}_{2} - \epsilon_{\rm NL, Y} \hat{\bm{q}}_{2}\right)}
\notag \\
& \times
\Biggl[ 1 + b^{\rm L}_{\rm X}b^{\rm L}_{\rm Y}\xi_{\rm L}(q)
-\frac{1}{2}k_{1i}k_{1j}A_{ij}(\hat{\bm{q}}_{1})
-\frac{1}{2}k_{2i}k_{2j}A_{ij}(\hat{\bm{q}}_{2})
\notag \\
&
- k_{1i}k_{2j}B_{ij}(\bm{q}_{1},\bm{q}_{2})
+ {\rm i} k_{2i} \left( b^{\rm L}_{\rm X}U_{i}(\bm{q}_{1},\bm{q}_{2})- b^{\rm L}_{\rm Y}y_{i}(\hat{\bm{q}}_{2}) \right)
\notag \\
&
+ {\rm i} k_{1i} \left( b^{\rm L}_{\rm Y}U_{i}(\bm{q}_{2},\bm{q}_{1}) - b^{\rm L}_{\rm X} y_{i}(\hat{\bm{q}}_{1}) \right)
\Biggr] ~.
\label{eq: DXDY linear}
\end{align}
The linearized expression of $\RX$ (or $\RY$) in Eq.~(\ref{eq: RX full app}) is given as
\begin{align}
\RX(\bm{s}_{1})
=
\int{\rm d}^{3}q\; 
\int&\frac{{\rm d}^{3}k}{(2\pi)^{3}}\;
{\rm e}^{{\rm i}\bm{k}\cdot \left( \bm{s}_{1} - \bm{q} - \epsilon_{\rm NL, X} \hat{\bm{q}}\right)}
\notag \\
& \times \left( 1 - {\rm i}b^{\rm L}_{\rm X}k_{i}y_{i} -\frac{1}{2}k_{i}k_{j}A_{ij}\right)
~.
\label{eq: RX linear}
\end{align}

By performing integration by parts, $\DD$ is rewritten as
\begin{align}
& \DD(\bm{s}_{1},\bm{s}_{2}) \notag \\
&=
J_{\rm X}(s_{1}) J_{\rm Y}(s_{2})
\notag \\
&
+
J_{\rm X}(s_{1})
\Biggl[
\frac{\partial}{\partial s_{2i}}
\Bigl\{ b^{\rm L}_{\rm Y} J_{\rm Y}(s_{2}) y_{i}(\hat{\bm{s}}_{2})\Bigr\}
+ \frac{\partial}{\partial s_{2i}}\frac{\partial}{\partial s_{2j}}
\Bigl\{ J_{\rm Y}(s_{2}) \frac{1}{2}A_{ij}(\hat{\bm{s}}_{2})\Bigr\}
\Biggr]
\notag \\
&
+
J_{\rm Y}(s_{2})
\Biggl[
\frac{\partial}{\partial s_{1i}}
\Bigl\{ b^{\rm L}_{\rm X} J_{\rm X}(s_{1}) y_{i}(\hat{\bm{s}}_{1})\Bigr\}
+ \frac{\partial}{\partial s_{1i}}\frac{\partial}{\partial s_{1j}}
\Bigl\{ J_{\rm X}(s_{1}) \frac{1}{2}A_{ij}(\hat{\bm{s}}_{1})\Bigr\}
\Biggr]
\notag \\
&
+ \int{\rm d}^{3}q_{1}
\int{\rm d}^{3}q_{2}
\int\frac{{\rm d}^{3}k_{1}}{(2\pi)^{3}}
\int\frac{{\rm d}^{3}k_{2}}{(2\pi)^{3}}
\notag \\
&\times 
{\rm e}^{{\rm i}\bm{k}_{1}\cdot \left( \bm{s}_{1} - \bm{q}_{1} - \epsilon_{\rm NL, X} \hat{\bm{q}}_{1}\right)
+{\rm i} \bm{k}_{2}\cdot \left( \bm{s}_{2} - \bm{q}_{2} - \epsilon_{\rm NL, Y} \hat{\bm{q}}_{2}\right)}
\Bigl[
b_{\rm X}^{\rm L}b_{\rm Y}^{\rm L}\,\xi_{\rm L}(q)
\notag \\
&
-k_{1,i}k_{2,j}\,B_{ij}(\bm{q}_1,\bm{q}_2)
+ {\rm i} k_{2i} b^{\rm L}_{\rm X}U_{i}(\bm{q}_{1},\bm{q}_{2}) 
+ {\rm i} k_{1i} b^{\rm L}_{\rm Y}U_{i}(\bm{q}_{2},\bm{q}_{1})
\Bigr]
~, \label{eq: DXDY linear part int}
\end{align}
and $\RX$ (or $\RY$) is rewritten as
\begin{align}
\RX(\bm{s}_{1})
=
J_{\rm X}(s_{1})
&+ \frac{\partial}{\partial s_{1i}}
\Bigl\{ b^{\rm L}_{\rm X} J_{\rm X}(s_{1}) y_{i}(\hat{\bm{s}}_{1})\Bigr\}
\notag \\
&
+ \frac{\partial}{\partial s_{1i}}\frac{\partial}{\partial s_{1j}}
\Bigl\{ J_{\rm X}(s_{1}) \frac{1}{2} A_{ij}(\hat{\bm{s}}_{1})\Bigr\}
~,
\label{eq: RX linear part int}
\end{align}
where we define the Jacobians $J_{\rm X}(s_{1})$ and $J_{\rm Y}(s_{2})$ of the transformation: $\tilde{\bm{q}}_{1} = \bm{q}_{1} + \epsilon_{\rm NL, X}\hat{\bm{q}}$ and $\tilde{\bm{q}}_{2} = \bm{q}_{2} + \epsilon_{\rm NL, Y} \hat{\bm{q}}_{2}$, respectively.
The explicit form of the Jacobians is given as
\begin{align}
J_{\rm X}(\tilde{q}_{1}) &\equiv \det{\left| \frac{\partial q_{1i}}{\partial \tilde{q}_{1j}}\right|} = \left( 1 + \frac{\epsilon_{\rm NL, X}}{\tilde{q}_{1}}\right)^{2}
~,\\
J_{\rm Y}(\tilde{q}_{2}) &\equiv \det{\left| \frac{\partial q_{2i}}{\partial \tilde{q}_{2j}}\right|} = \left( 1 + \frac{\epsilon_{\rm NL, Y}}{\tilde{q}_{2}}\right)^{2} ~.
\end{align}
Substituting Eqs.~(\ref{eq: DXDY linear part int}) and (\ref{eq: RX linear part int}) into the definition of the correlation function~(\ref{eq: xi def app}), almost all terms cancel except for the last three lines in Eq.~(\ref{eq: DXDY linear part int}).

Finally we obtain the linearized correlation function as
\begin{align}
& \xi_{\rm XY}^{\rm(S)}(\bm{s}_1,\bm{s}_2) \notag \\
&= 
\left( J_{\rm X}(s_{1})J_{\rm Y}(s_{2}) \right)^{-1}
\int{\rm d}^{3}q_{1}
\int{\rm d}^{3}q_{2}
\int\frac{{\rm d}^{3}k_{1}}{(2\pi)^{3}}
\int\frac{{\rm d}^{3}k_{2}}{(2\pi)^{3}}
\notag \\
&\times 
{\rm e}^{{\rm i}\bm{k}_{1}\cdot \left( \bm{s}_{1} - \bm{q}_{1} - \epsilon_{\rm NL, X} \hat{\bm{q}}_{1}\right)
+{\rm i} \bm{k}_{2}\cdot \left( \bm{s}_{2} - \bm{q}_{2} - \epsilon_{\rm NL, Y} \hat{\bm{q}}_{2}\right)}
\Bigl[
b_{\rm X}^{\rm L}b_{\rm Y}^{\rm L}\,\xi_{\rm L}(q)
\notag \\
&
-k_{1,i}k_{2,j}\,B_{ij}(\bm{q}_1,\bm{q}_2)
+ {\rm i} k_{2i} b^{\rm L}_{\rm X}U_{i}(\bm{q}_{1},\bm{q}_{2}) 
+ {\rm i} k_{1i} b^{\rm L}_{\rm Y}U_{i}(\bm{q}_{2},\bm{q}_{1})
\Bigr]
\notag \\
&= 
\left( J_{\rm X}(s_{1})J_{\rm Y}(s_{2}) \right)^{-1}
\Biggl[
J_{\rm X}(s_{1})J_{\rm Y}(s_{2})
b_{\rm X}^{\rm L}b_{\rm Y}^{\rm L}\,\xi_{\rm L}(q)
\notag \\
&
+\frac{\partial}{\partial s_{1i}}\frac{\partial}{\partial s_{2j}}
\Bigl\{ J_{\rm X}(s_{1})J_{\rm Y}(s_{2}) B_{ij}(\bm{q}_{1},\bm{q}_{2}) \Bigr\}
\notag \\
&
+\frac{\partial}{\partial s_{2i}}
\Bigl\{ b^{\rm L}_{\rm X} J_{\rm X}(s_{1})J_{\rm Y}(s_{2}) U_{i}(\bm{q}_{1},\bm{q}_{2}) \Bigr\}
\notag \\
&
+\frac{\partial}{\partial s_{1i}}
\Bigl\{ b^{\rm L}_{\rm Y} J_{\rm X}(s_{1})J_{\rm Y}(s_{2}) U_{i}(\bm{q}_{2},\bm{q}_{1}) \Bigr\}
\Biggr]~,
\end{align}
where, in the second equality, the arguments of the functions $\xi_{\rm L}$, $B_{ij}$, and $U_{i}$, are related to the redshift space $\bm{s}_{1}$ and $\bm{s}_{2}$ by $\bm{q}_{1} = \bm{s}_{1} - \epsilon_{\rm NL, X} \hat{\bm{s}}_{1}$ and $\bm{q}_{2} = \bm{s}_{2} - \epsilon_{\rm NL, Y} \hat{\bm{s}}_{2}$, respectively.

Neglecting the higher order terms, $O\left( \epsilon^{2}_{\rm NL,X/Y} \right)$ and $O\left( \epsilon_{\rm NL, X/Y} P_{\rm L}(k) \right)$, we obtain the cross-correlation function expressed as a function of $\bm{s}_{1}$ and $\bm{s}_{2}$:
\begin{align}
\xi_{\rm XY}^{\rm(S)}(\bm{s}_1,\bm{s}_2)
&= 
b_{\rm X}^{\rm L}b_{\rm Y}^{\rm L}\,\xi_{\rm L}(s)
+\frac{\partial}{\partial s_{1i}}
\frac{\partial}{\partial s_{2j}}
\,B_{ij}(\bm{s}_{1},\bm{s}_{2})
\notag \\
&
+ b^{\rm L}_{\rm X} \frac{\partial}{\partial s_{2i}} U_{i}(\bm{s}_{1},\bm{s}_{2})
+ b^{\rm L}_{\rm Y} \frac{\partial}{\partial s_{1i}} U_{i}(\bm{s}_{2},\bm{s}_{1})
~. \label{eq: corr linear s1 s2}
\end{align}

To further reduce Eq.~(\ref{eq: corr linear s1 s2}), we evaluate the spacial derivatives of the $B_{ij}$ and $U_{i}$ as
\begin{align}
\frac{\partial}{\partial s_{1i}} & U_{i}(\bm{s}_{2},\bm{s}_{1})
=
\int\frac{{\rm d}^{3}k}{(2\pi)^{3}}\; {\rm e}^{{\rm i}\bm{k}\cdot\bm{s}}P_{L}(k)
\notag \\
&\times 
\Biggl[
1 + f \mu^{2}_{1} + {\rm i} 2 f\frac{\mu_{1}}{ks_{1}}
- \frac{3}{2}\frac{\Omega_{\rm m0}H^{2}_{0}}{a^{2}H}\frac{1}{k} \left( \frac{2}{ks_{1}} - i\mu_{1} \right)
\Biggr] ~, \label{eq: partial U2}
\end{align}
\begin{align}
\frac{\partial}{\partial s_{2i}} & U_{i}(\bm{q}_{1},\bm{q}_{2})
=
\int\frac{{\rm d}^{3}k}{(2\pi)^{3}}\; {\rm e}^{{\rm i}\bm{k}\cdot\bm{s}}P_{L}(k)
\notag \\
&\times 
\Biggl[
1 + f\mu^{2}_{2} - {\rm i} 2 f\frac{\mu_{2}}{k s_{2}}
- \frac{3}{2}\frac{\Omega_{\rm m0}H^{2}_{0}}{a^{2}H} \frac{1}{k} \left( \frac{2}{k s_{2}} + i\mu_{2} \right)
\Biggr] ~, \label{eq: partial U1}
\end{align}
\begin{align}
\frac{\partial}{\partial s_{1a}}&
\frac{\partial}{\partial s_{2b}} B_{ab}
=
\int\frac{{\rm d}^{3}k}{(2\pi)^{3}}\;
{\rm e}^{{\rm i}\bm{k}\cdot\bm{s}}P_{L}(k)
\notag \\
&\times 
\Biggl[
1 + f\mu^{2}_{1} + {\rm i} 2 f \frac{\mu_{1}}{ks_{1}}
- \frac{3}{2}\frac{\Omega_{\rm m0}H^{2}_{0}}{a^{2}H} \frac{1}{k}\left( \frac{2}{ks_{1}} - {\rm i} \mu_{1}\right)
\Biggr]
\notag \\
&\times 
\Biggl[
1 + f\mu^{2}_{2} - {\rm i} 2 f \frac{\mu_{2}}{ks_{2}}
- \frac{3}{2}\frac{\Omega_{\rm m0}H^{2}_{0}}{a^{2}H} \frac{1}{k}\left( \frac{2}{ks_{2}} + {\rm i} \mu_{2}\right)
\Biggr] ~, \label{eq: partial B}
\end{align}
where the directional cosine $\mu_{1,2}$ is defined by 
$\mu_{1,2} = \hat{\bm{k}}\cdot\hat{\bm{s}}_{1,2}$ in the above expressions.

Collecting all the contributions to the correlation function, we finally obtain
\begin{align}
\xi_{\rm XY}^{\rm(S)}& (\bm{s}_1,\bm{s}_2) =
\int\frac{{\rm d}^{3}k}{(2\pi)^{3}} {\rm e}^{{\rm i}\bm{k}\cdot\bm{s}} P_{L}(k)
\notag \\
& \times \Biggl[
b_{\rm X} + f\mu^{2}_{1} + {\rm i} 2 f \frac{\mu_{1}}{ks_{1}}
- \frac{3}{2}\frac{\Omega_{\rm m0}H^{2}_{0}}{a^{2}H} \frac{1}{k}\left( \frac{2}{ks_{1}} - {\rm i} \mu_{1}\right)
\Biggr]
\notag \\
&\times
\Biggl[
b_{\rm Y} + f\mu^{2}_{2} - {\rm i} 2 f \frac{\mu_{2}}{ks_{2}}
- \frac{3}{2}\frac{\Omega_{\rm m0}H^{2}_{0}}{a^{2}H} \frac{1}{k}\left( \frac{2}{ks_{2}} + {\rm i} \mu_{2}\right)
\Biggr] ~. \label{eq: xi linear theory}
\end{align}
We introduce the Eulerian linear bias given by $b_{\rm X, Y} = 1 + b^{\rm L}_{\rm X,Y}$.
The integrand of this expression coincides with the linear theory result presented in \citet{2009JCAP...11..026M}, except for the selection function terms that is proportional to $2/s_{1}$ and $2/s_{2}$.
Note that dropping the last terms in the second and third lines, which is relevant to the gravitational redshift effect, the expression coincides with the linear theory expression without taking the distant-observer limit~(e.g., \citealt{2008MNRAS.389..292P,2015MNRAS.447.1789Y,2016JCAP...01..048R,2020MNRAS.491.4162T}).

\section{Analytical expressions for the linear cross-correlation function}
\label{appendix: linear theory}

The analytical expressions for the linearized cross-correlation function with wide-angle effects have  been already presented in Appendix~D of \citet{2020MNRAS.491.4162T}.
In this appendix, apart from the formula for the standard Doppler contribution with wide-angle effects, we present the analytical expressions for the new contribution coming from the gravitational redshift effect.

Starting from the expression at Eq.~(\ref{eq: xi linear theory}), we first decompose the linear cross-correlation function into two pieces: $\xi^{({\rm S})}_{\rm XY} = \xi^{({\rm S})}_{\rm XY, Doppler} + \xi^{({\rm S})}_{\rm XY, rel}$, where $\xi^{({\rm S})}_{\rm XY, Doppler}$ and $\xi^{({\rm S})}_{\rm XY, grav}$, respectively, stand for the standard Doppler and gravitational redshift effects.
Collecting the gravitational redshift contribution to Eq.~(\ref{eq: xi linear theory}), the expression of $\xi^{({\rm S})}_{\rm XY, grav}$ is written as
\begin{align}
&\xi^{({\rm S})}_{\rm XY, grav}(\bm{s}_1,\bm{s}_2) =
\int\frac{{\rm d}^{3}k}{(2\pi)^{3}} {\rm e}^{{\rm i}\bm{k}\cdot\bm{s}} P_{L}(k)
\notag \\
& \times
\Biggl[
\Biggl(
b_{\rm X} + f\mu^{2}_{1} + {\rm i} 2 f \frac{\mu_{1}}{ks_{1}}
\Biggr)
\Biggl(
- \frac{3}{2}\frac{\Omega_{\rm m0}H^{2}_{0}}{a^{2}H} \frac{1}{k}\left( \frac{2}{ks_{2}} + {\rm i} \mu_{2}\right)
\Biggr)
\notag \\
& +
\Biggl(
b_{\rm Y} + f\mu^{2}_{2} - {\rm i} 2 f \frac{\mu_{2}}{ks_{2}}
\Biggr)
\Biggl(
- \frac{3}{2}\frac{\Omega_{\rm m0}H^{2}_{0}}{a^{2}H} \frac{1}{k}\left( \frac{2}{ks_{1}} - {\rm i} \mu_{1}\right)
\Biggr)
\notag \\
& +
\Biggl(
- \frac{3}{2}\frac{\Omega_{\rm m0}H^{2}_{0}}{a^{2}H} \frac{1}{k}\left( \frac{2}{ks_{1}} - {\rm i} \mu_{1}\right)
\Biggr)
\Biggl(
- \frac{3}{2}\frac{\Omega_{\rm m0}H^{2}_{0}}{a^{2}H} \frac{1}{k}\left( \frac{2}{ks_{2}} + {\rm i} \mu_{2}\right)
\Biggr)
\Biggr] ~. \label{eq: xi linear theory new}
\end{align}
In the following subsections, we will focus on the gravitational redshift contribution to the cross-correlation function given in Eq.~(\ref{eq: xi linear theory new}).

\subsection{Expansion form of linear cross-correlation function}

In \citet{2004ApJ...614...51S,2008MNRAS.389..292P}, the linear-order correlation function with wide-angle effects is expanded in terms of the tripolar spherical harmonics. 
\citet{2020MNRAS.491.4162T} extend their treatment to the linear-order correlation function between different biased objects.
In this subsection, following \citet{2004ApJ...614...51S,2008MNRAS.389..292P,2011PhRvD..84f3505B}, we further extend the treatment to include the gravitational redshift contribution to the linear-order cross-correlation function.

We first define the tripolar spherical harmonics by
\begin{align}
S_{\ell_{1},\ell_{2},\ell}(\hat{\bm{s}}_{1},\hat{\bm{s}}_{2},\hat{\bm{s}}) &= \sum_{m_{1},m_{2},m}
\left(
\begin{array}{ccc}
\ell_{1} & \ell_{2} & \ell \\
m_{1} & m_{2} & m
\end{array}
\right)\notag \\
&\times 
C^{*}_{\ell_{1},m_{1}}(\hat{\bm{s}}_{1})
C^{*}_{\ell_{2},m_{2}}(\hat{\bm{s}}_{2})
C^{*}_{\ell,m}(\hat{\bm{s}}) ~,
\end{align}
where the matrix is a Wigner-$3j$ symbol.
The function, $C_{\ell, m}(\hat{\bm{s}}) = \sqrt{4\pi/(2\ell +1)}Y_{\ell,m}(\hat{\bm{s}})$, is the normalized spherical harmonics.

The tripolar spherical harmonics characterize the angular dependence of cross-correlation function, and then, we separate the dependence of the distance and separation from their angular dependence in the cross-correlation function as
\begin{equation}
\xi^{(\rm S)}_{\rm XY, grav}(\bm{s}_{1},\bm{s}_{2}) = \sum_{\ell_{1},\ell_{2},\ell} b_{\ell_{1}\ell_{2}\ell}(s_{1},s_{2},s) S_{\ell_{1},\ell_{2},\ell}(\hat{\bm{s}}_{1},\hat{\bm{s}}_{2},\hat{\bm{s}}) ~, \label{eq: tripolar expansion}
\end{equation}
where the coefficient, $b_{\ell_{1}\ell_{2}\ell}$, is given as the function of $s_{1} = |\bm{s}_{1}|$, $s_{2} = |\bm{s}_{2}|$, and $s = |\bm{s}_{2}-\bm{s}_{1}|$.
The non-vanishing coefficients are summarized as follows:
\begin{align}
b_{000} & =
- \Biggl\{\frac{2}{s_{1}}\left( b_{\rm Y}+\frac{f}{3} \right)
+\frac{2}{s_{2}}\left( b_{\rm X}+\frac{f}{3} \right) \Biggr\} \eta_{0}^{2}(s)
+ \frac{4}{s_{1}s_{2}} \lambda_{0}^{2}(s) ~,
\\
b_{101} &=
\sqrt{3} \left( b_{\rm Y} + \frac{f}{3} \right) \eta_{1}^{3}(s) 
- \frac{2\sqrt{3}}{s_{2}} \Biggl\{ \lambda_{1}^{3}(s) + \frac{2f}{s_{1}} \eta_{1}^{1}(s)\Biggr\} ~,
\\
b_{011} & = -\sqrt{3}\left( b_{\rm X}+\frac{f}{3}\right) \eta_{1}^{3}(s)
+ \frac{2\sqrt{3}}{s_{1}}\Biggl\{ \lambda_{1}^{3}(s) + \frac{2f}{s_{2}} \eta_{1}^{1}(s)\Biggr\} ~,
\\
b_{110} &= -\frac{1}{\sqrt{3}} \Biggl\{ 2\left( \frac{f}{s_{1}} + \frac{f}{s_{2}}\right) \eta_{0}^{2}(s) + \lambda_{0}^{4}(s)\Biggr\} ~,
\\
b_{112} &= -\sqrt{\frac{10}{3}} \Biggl\{ 2\left( \frac{f}{s_{1}} + \frac{f}{s_{2}}\right) \eta_{2}^{2}(s) + \lambda_{2}^{4}(s) \Biggr\} ~,
\\
b_{022} &= \frac{4\sqrt{5} f}{3 s_{1}} \eta_{2}^{2}(s) ~,
\\
b_{202} & = \frac{4\sqrt{5} f}{3 s_{2}} \eta_{2}^{2}(s) ~,
\\
b_{211} &= -b_{121} = 2\sqrt{\frac{2}{15}}f \eta_{1}^{3}(s) ~,
\\
b_{213} &= -b_{123} = 2\sqrt{\frac{7}{15}}f \eta_{3}^{3}(s) ~,
\end{align}
where we define $\eta^{n}_{\ell}(s)$ and $\lambda^{n}_{\ell}(s)$ by
\begin{align}
\eta^{n}_{\ell}(s) &= \int\frac{{\rm d}k}{2\pi^{2}} k^{n}j_{\ell}(ks) P_{\delta \epsilon}(k) ~, \label{eq: def eta ell m}\\
\lambda^{n}_{\ell}(s) &= \int\frac{{\rm d}k}{2\pi^{2}} k^{n}j_{\ell}(ks) P_{\epsilon}(k) ~.
\end{align}
Using Eqs.~(\ref{eq: potential app}) and (\ref{eq: Fourier Poisson}), the power spectra $P_{\delta\epsilon}(k)$ and $P_{\epsilon}(k)$ are given in terms of the linear matter power spectrum $P_{\rm L}(k)$ as
\begin{align}
P_{\delta\epsilon}(k) &= \left( \frac{3\Omega_{\rm m0}H^{2}_{0}}{2aH} \frac{1}{k^{2}}\right) P_{\rm L}(k) ~,\\
P_{\epsilon}(k) &= \left( \frac{3\Omega_{\rm m0}H^{2}_{0}}{2aH} \frac{1}{k^{2}}\right)^{2} P_{\rm L}(k) ~.
\end{align}
Note that, if we use the another definition of the separation vector given by $\bm{s} = \bm{s}_{1}-\bm{s}_{2}$, the sign for the terms involving either of factor $1/s_{1}$ or $1/s_{2}$ are flipped.

Choosing a specific coordinate system, we further simplify Eq.~(\ref{eq: tripolar expansion}).
Here we use the same coordinate system defined in~\citet{2008MNRAS.389..292P,2014PhRvD..89h3535B,2020MNRAS.491.4162T}:
\begin{equation}
\hat{\bm{s}}_{1} = (\cos{\phi_{1}},\sin{\phi_{1}},0) ~,~~~
\hat{\bm{s}}_{2} = (\cos{\phi_{2}},\sin{\phi_{2}},0) ~.
\end{equation}
This coordinate system is chosen so that the triangle formed with the position vectors $\bm{s}_{1}$ and $\bm{s}_{2}$ is confined on the $x$-$y$ plane, and the separation vector $\bm{s} = \bm{s}_{2} -\bm{s}_{1}$ is parallel to the $x$-axis.
Then, the cross-correlation function is expanded as
\begin{align}
\xi^{(\rm S)}_{\rm XY, grav}(\bm{s}_{1},\bm{s}_{2})
=& \sum_{m,n}\Bigl[ a_{mn}\cos{(m\phi_{1})}\cos{(n\phi_{2})}\notag \\
& + b_{mn}\sin{(m\phi_{1})}\sin{(n\phi_{2})}\Bigr] ~,
\end{align}
where $a_{mn}$ and $b_{mn}$ are given as the function of $s_{1}$, $s_{2}$, and $s$.
The non-vanishing coefficients are summarized as follows:
\begin{align}
a_{00} & = 
-2\Biggl\{ 
\frac{1}{s_{1}}\left( b_{\rm Y}+\frac{f}{3}\right) +
\frac{1}{s_{2}}\left( b_{\rm X}+\frac{f}{3}\right) 
 \Biggr\}\eta_{0}^{2}(s)
\notag \\
& + f \left( \frac{1}{s_{1}}+\frac{1}{s_{2}} \right) \eta_{2}^{2}(s)
+ \frac{4}{s_{1}s_{2}} \lambda_{0}^{2}(s) ~,\\
a_{01} & = \left( b_{\rm X} +\frac{2 f}{5} \right) \eta_{1}^{3}(s) - \frac{f}{10} \eta_{3}^{3}(s) 
- \frac{4\,f}{s_{1}s_{2}} \eta_{1}^{1}(s) - \frac{2}{s_{1}} \lambda_{1}^{3}(s) ~, \\
a_{10} & = - \left( b_{\rm Y} + \frac{2f}{5} \right) \eta_{1}^{3}(s) + \frac{f}{10} \eta_{3}^{3}(s)
+ \frac{4\,f}{s_{1}s_{2}} \eta_{1}^{1}(s) + \frac{2}{s_{2}} \lambda_{1}^{3}(s) ~, \\
a_{11} & = \frac{2 f}{3} \left( \frac{1}{s_{1}}+\frac{1}{s_{2}}\right) \left( \eta_{0}^{2}(s)-2 \eta_{2}^{2}(s) \right)
+\frac{1}{3}\left( \lambda_{0}^{4}(s)-2\lambda_{2}^{4}(s) \right) ~, \\
a_{02} & = \frac{f}{s_{1}} \eta_{2}^{2}(s) ~,\\
a_{20} & = \frac{f}{s_{2}} \eta_{2}^{2}(s) ~,\\
a_{12} & = - a_{21} = \frac{f}{5}\left( -\eta_{1}^{3}(s) + \frac{3}{2} \eta_{3}^{3}(s)\right) ~, \\
b_{11} & = \frac{2 f}{3}\left( \frac{1}{s_{1}}+\frac{1}{s_{2}}\right) \left( \eta_{0}^{2}(s)+\eta_2^2(s) \right)
+\frac{1}{3} \left( \lambda_{0}^{4}(s)+\lambda_{2}^{4}(s) \right) ~,\\
b_{12} & = -b_{21} = - \frac{f}{5} \left( \eta_{1}^{3}(s)+\eta_{3}^{3}(s) \right) ~.
\end{align}
These expressions are suited for numerically computing the cross-correlation function.

\subsection{Wide-angle corrections for cross-correlation function with relativistic effects}
\label{appendix: wide-angle corrections}

The multipole moment of the correlation function generally depends on both the separation $s$ and the line-of-sight distance $d$.
We then expand the multipole moment in powers of $(s/d)$ as the wide-angle corrections:
\begin{align}
\xi^{(\rm S)}_{{\rm XY}}(s,d,\mu) &= \sum_{\ell}\xi^{(\rm S)}_{{\rm XY}, \ell}(s,d) \mathcal{P}_{\ell}(\mu) \notag \\
&= \sum_{\ell}\sum_{n} \left( \frac{s}{d}\right)^{n}\xi^{(\rm S)}_{\ell,n}(s) \mathcal{P}_{\ell}(\mu) ~. \label{eq: wa expansion app}
\end{align}
In this subsection, we derive the expression for $\xi^{({\rm S})}_{\ell,n}(s)$ up to the next-to-leading order adopting the mid-point line-of-sight.

The non-vanishing multipoles at leading order up to $\ell = 4$ are summarized as follows:
\begin{align}
\xi^{(\rm S)}_{0,0} &= \frac{1}{3}\lambda^{4}_{0}(s) ~,\\
\xi^{(\rm S)}_{1,0} &= (b_{\rm X} - b_{\rm Y}) \eta^{3}_{1}(s) ~, \label{eq: xi 1 0} \\
\xi^{(\rm S)}_{2,0} &= -\frac{2}{3}\lambda^{4}_{2}(s) ~, \\
\xi^{(\rm S)}_{3,2} &= \frac{3}{20}(b_{\rm X}-b_{\rm Y})\eta^{3}_{1}(s) ~, \label{eq: xi 3 2} \\
\xi^{(\rm S)}_{4,1} &= -\frac{8}{35} f \eta_{3}^{3}(s) ~,
\end{align}
and the non-vanishing multipoles at next-to-leading order are summarized as follows:
\begin{align}
\xi^{(\rm S)}_{0,1} &= \frac{1}{3} \left( b_{\rm X} + b_{\rm Y} -\frac{2}{5}f \right)\eta_{1}^{3}(s) - \frac{2}{s}(b_{\rm X}+b_{\rm Y})\eta_{0}^{2}(s) ~, \\
\xi^{(\rm S)}_{1,2} &= (b_{\rm X} - b_{\rm Y}) \left( -\frac{3}{20} \eta^{3}_{1}(s)
+ \frac{1}{s} \eta^{2}_{0}(s) \right)~,\\
\xi^{(\rm S)}_{2,1} &= - \frac{1}{3} \left( b_{\rm X}+b_{\rm Y} - \frac{2}{5}f \right)\eta_{1}^{3}(s) + \frac{8}{35}f \eta_{3}^{3}(s) ~,\\
\xi^{(\rm S)}_{3,4} &= (b_{\rm X}-b_{\rm Y})\left( - \frac{5}{144}\eta^{3}_{1}(s) + \frac{1}{4s}\eta^{2}_{0}(s) \right) ~,\\
\xi^{(\rm S)}_{4,2} &= -\frac{4}{35}\lambda^{4}_{2}(s) ~.
\end{align}
The odd multipole moments are proportional to the bias difference $b_{\rm X}-b_{\rm Y}$.

\section{Analytical model of non-perturbative halo potential}
\label{appendix: halo potential}

In this appendix, based on the NFW profile and its gravitational potential~\citep{1996ApJ...462..563N}, we present an analytical model for the non-perturbative halo potential defined in Eq.~(\ref{eq: non-linear potential}) 

The NFW profile is known to quantitatively describe the halo density profiles in cosmological $N$-body simulations, and it is given by
\begin{equation}
\rho_{\rm NFW}(r, z, M) =
\frac{\rho_{\rm s}(z, M)}
{\left( r/r_{\rm s}(z, M) \right)
\left\{ 1+\left( r/r_{\rm s}(z, M) \right) \right\}^{2}} ~,
\label{eq: NFW profile}
\end{equation}
where $r$, $z$, and $M$ are radius from the centre of halo, redshift, and halo mass, respectively. The overdensity, $\rho_{\rm s}(z,M)$, and the scale radius, $r_{\rm s}(z,M)$, are related to the concentration parameter $c_{\rm vir}$ through
\begin{align}
\rho_{\rm s}(z, M) &= \frac{\Delta_{\rm vir}(z)\rho_{\rm m0}}{3}c^{3}_{\rm vir}(M, z) 
\notag \\
&\times 
\left[ \ln{\left( 1 +c_{\rm vir}(z, M) \right)} - \frac{c_{\rm vir}(z, M)}{1+c_{\rm vir}(z, M)} \right]^{-1}
~, \\
r_{\rm s}(z, M) &= \frac{r_{\rm vir}(z, M)}{c_{\rm vir}(z, M)} ~, 
\end{align}
with the radius $r_{\rm vir}$ and $\Delta_{\rm vir}$ being the virial radius and virial overdensity, respectively given by  \citep{1998ApJ...495...80B,2001MNRAS.321..559B}
\begin{align}
r_{\rm vir}(z, M) &= \left( \frac{3M}{4\pi\Delta_{\rm vir}(z)\rho_{\rm m0}} \right)^{\frac{1}{3}} ~,
\end{align}
\begin{align}
\Delta_{\rm vir}(z) &= \frac{18\pi^{2} + 82\left( \Omega_{\rm m}(z) -1\right) - 39\left( \Omega_{\rm m}(z) -1\right)^{2}}{\Omega_{\rm m}(z)}~,\\
\Omega_{\rm m}(z) &= \frac{(1+z)^{3}\Omega_{\rm m0}}{(1+z)^{3}\Omega_{\rm m0} + \Omega_{\Lambda 0}} ~.
\end{align}
The concentration parameter $c_{\rm vir}(z, M)$ is known to weakly depend on the halo mass, and we use the following fitting form~\citep{2001MNRAS.321..559B,2002PhR...372....1C}: 
\begin{equation}
c_{\rm vir}(z, M) = \frac{9}{1+z}\left( \frac{M}{M_{*}(z)}\right)^{-0.13} ~,
\end{equation}
where $M_{*}(z)$ stands for the characteristic mass scale defined by $\sigma_{M}(M_{*})D_{+}(z) = \delta_{\rm crit}$. The quantity $\delta_{\rm crit}$ is the critical over-density of the spherical collapse model, and $\sigma_{M}$ is the root-mean square amplitude of the matter density fluctuations smoothed with top-hat filter of the radius $R = \left( 3M/(4\pi \overline{\rho})\right)^{1/3}$ with $\overline{\rho}$ being the mean mass density.

Solving the Poisson equation, the gravitational potential of the NFW profile at Eq.~(\ref{eq: NFW profile}) is obtained under the boundary condition $\phi_{\rm NFW}\to0$ at $r\to\infty$. We have
\begin{align}
\phi_{\rm NFW}(r, z, M) &= -4\pi G (1+z) \rho_{\rm s}(z, M) r^{2}_{\rm s}(z, M)
\notag \\
& \times
\left(\frac{r}{r_{\rm s}(z, M)}\right)^{-1}\ln{\left( 1 + \frac{r}{r_{\rm s}(z, M)}\right)} ~. \label{eq: phi NFW(r)}
\end{align}
Taking the limit $r\to 0$, the gravitational potential at the centre of the halo, which we denote by $\phi_{\rm NFW, 0}$, is estimated to be 
\begin{equation}
\phi_{\rm NFW, 0}(z, M) = -4\pi G (1+z) \rho_{\rm s}(z, M) r^{2}_{\rm s}(z, M) ~.
\label{eq: phi NFW origin}
\end{equation}

In comparing the model predictions with the measured halo potentials in $N$-body simulations, one must properly take account of the finite bin width of the halo mass scale in the halo catalogues. That is, the prediction of the potential has to be averaged over the halo mass, weighting with the halo mass function $n(M)$:
\begin{align}
\overline{\phi}_{\rm NFW}(z,M) &= \frac{\int^{M_{\rm max}}_{M_{\rm min}}\; \phi_{\rm NFW,0}(z, M)n(M) \;{\rm d}M}
{\int^{M_{\rm max}}_{M_{\rm min}}\; n(M) \;{\rm d}M} ~, \label{eq: phibar NFW}
\end{align}
where the upper and lower limit of the integral are taken to be $M_{\rm max}=2 N m_{\rm DM}$ and $M_{\rm min}=N m_{\rm DM}$, so as to be the same mass range as in the halo catalogues used in our analysis (see Table~\ref{tab: population}), where $m_{\rm DM}$ is the mass of the dark matter particle. In Sec.~\ref{subsec:modelling_potential}, the predictions based on Eq.~(\ref{eq: phibar NFW}) is compared in detail with the measured non-linear potentials.

\section{Impact of halo identification on the gravitational redshift effect}
\label{appendix: b = 0.1}

In this appendix, we perform the same analysis as presented in Sec.~\ref{sec: halo potential sim}, but with a different criterion for halo definition, and compare the measured non-linear potential with the analytical prediction based on NFW profile.
The $N$-body simulation used in this appendix contains $1024^{3}$ dark matter particles in a cosmological volume of $(656.25\, {\rm Mpc}/h)^{3}$.

In the main text, we used haloes identified with the friend-of-friend algorithm, adopting the linking length $b= 0.2$ times the mean inter-particle separation.
Here, we adopt the smaller linking length of $b=0.1$.
With a smaller linking length, the friend-of-friend algorithm selects more dense region for halos, and one expects that the shape of the identified haloes becomes spherically symmetric, and the center-of-mass position gets closer to the position of the deepest potential, similarly to the situation considered in the analytical model. Thus, the choice of smaller linking length can mitigate, to some extent, the discrepancy between the measured results and analytical prediction seen in Fig.\ref{fig: halo potential}.

\begin{figure}
\centering
\includegraphics[width=\columnwidth]{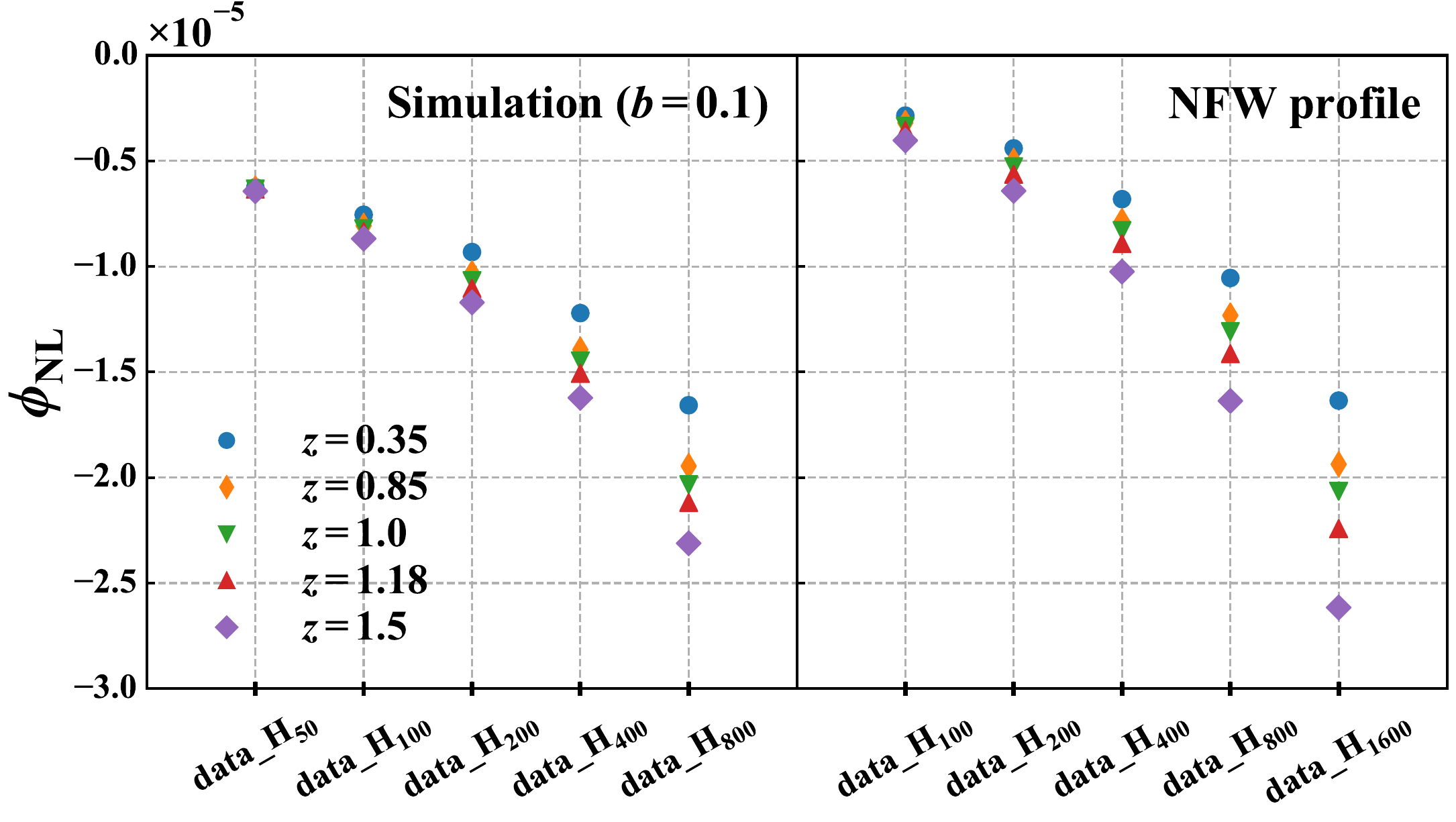}
\caption{
Same as Fig.~\ref{fig: halo potential} but for using the linking length $b=0.1$ times the mean inter-particle separation to detect haloes in $N$-body simulations.
Note that the redshifts are slightly different from Fig.~\ref{fig: halo potential}.
}
\label{fig: halo potential b=0.1}
\end{figure}

Fig.~\ref{fig: halo potential b=0.1} is the same plot as in Fig.~\ref{fig: halo potential}, but the results with the smaller linking length are shown.
Using the halo definition with $b=0.1$, an enclosed over-density is approximately $8$ times larger than that with $b=0.2$~\citep{2008ApJ...688..709T}.
Since the halo mass ratio between these two definitions is about factor 2 at $z\approx 0$~\citep{2014MNRAS.437.2328B}, we present \data{N} in simulations with  $b=0.1$ and \data{2N} in the theoretical prediction, for a consistent comparison.
Note that redshifts of the output results are slightly different from those used in Fig.~\ref{fig: halo potential}.
As we expected, the analytical predictions now get closer to the measured results, especially the massive halo populations.
On the other hand, there still remains the discrepancy for the light haloes.
As we mentioned in Sec.~\ref{sec: halo potential sim}, in simulations, the halo potential tends to be affected by the surrounding environment in general, and the measured potential can get deeper than that of the prediction based on the isolated model, especially for small haloes.
Therefore, the remaining discrepancy for the light haloes would be attributed to the effect by the surrounding environment.

\bsp	
\label{lastpage}
\end{document}